\documentclass[twocolumn,apj,numberedappendix,twocolappendix]{openjournal}

\usepackage[utf8]{inputenc}
\usepackage{txfonts}

\usepackage{amssymb}

\usepackage{graphicx}
\DeclareGraphicsExtensions{.bmp,.png,.jpg,.pdf}
\usepackage{multirow}
\usepackage{makecell}
\usepackage{float}

\usepackage{xcolor}
\definecolor{linkcolor}{rgb}{0.0,0.3,0.5}
\usepackage[table]{xcolor} 
\usepackage{soul}
\usepackage{color,colortbl}

\usepackage{textgreek}
\usepackage[autopunct=true,autostyle=false]{csquotes}

\usepackage{tensind}
\tensordelimiter{?}
\usepackage{verbatim}
\usepackage[normalem]{ulem}
\usepackage{orcidlink}
\usepackage{soul}
\usepackage{tabularx}
\newcommand{\orcidauthor}[3]{\author{\href{http://orcid.org/#1}{#2$^{#3}$}}}
\DeclareUnicodeCharacter{0306}{}

\newcommand{\sIivoiv}{Si~IV+O~IV] $\lambda\lambda 1400$}
\newcommand{\hzetahei}{H8+HeI $\lambda\lambda 3889$}
\newcommand{\mgiid}{Mg~II $\lambda\lambda 2795$, 2802}
\newcommand{\sIii}{Si~II $\lambda 1812$}
\newcommand{\sIiid}{Si~II $\lambda\lambda 1808$, 1817}
\newcommand{\ciiid}{C~III] $\lambda\lambda 1907$, 1909}
\newcommand{\civd}{C~IV $\lambda\lambda 1548$, 1551}
\newcommand{\oiiit}{O~III] $\lambda\lambda 1661$, 1666}

\newcommand{\oiiioptred}{[O~III] $\lambda 5007$}
\newcommand{\nivd}{N~IV] $\lambda\lambda 1483$, 1488}

\newcommand{\feivd}{[Fe~IV] $\lambda\lambda 2832$, 2835}
\newcommand{\mgii}{Mg~II $\lambda 2800$}
\newcommand{\ciii}{C~III] $\lambda 1908$}
\newcommand{\civ}{C~IV $\lambda 1549$}
\newcommand{\heii}{He~II $\lambda 1640$}
\newcommand{\oiii}{O~III] $\lambda 1663$}
\newcommand{\oii}{[O~II] $\lambda 3727$}
\newcommand{\niii}{N~III] $\lambda 1750$}
\newcommand{\niv}{N~IV] $\lambda 1488$}
\newcommand{\neiii}{[Ne~III] $\lambda 3868$}
\newcommand{\neiv}{[Ne~IV] $\lambda 2424$}
\newcommand{\feiv}{[Fe~IV] $\lambda 2833$}
\newcommand{\nev}{[Ne~V] $\lambda 3426$}
\newcommand{\oiiibowen}{O~III $\lambda 3133$}

\newcommand{\zzsun}{$Z/{\rm Z}_{\odot}$}
\newcommand{\ergs}{erg~s$^{-1}$~cm$^{-2}$}
\newcommand{\ergsA}{erg~s$^{-1}$~cm$^{-2}$ \AA$^{-1}$}

\newcommand{\pyneb}{\texttt{PyNeb}}

\newcommand{\LF}{\ensuremath{\mathcal{L}}}
\defcitealias{Castellano2024}{C24}

\usepackage{hyperref}
\hypersetup{
    unicode,
    colorlinks=true,
    linkcolor=linkcolor,
    citecolor=linkcolor,
    filecolor=linkcolor,
    urlcolor=linkcolor,
}

\shorttitle{The ionising sources and complex ISM of GHZ2}
\shortauthors{M. Castellano et al.}
\catcode`\'=12
\begin{document}
\title{Investigating ionising sources and the complex interstellar medium of GHZ2 at $z=12.3$\vspace{-1.5cm}}
\orcidauthor{0000-0001-9875-8263}{M. Castellano$^\dagger$}{1}
\orcidauthor{0000-0002-8951-4408}{L. Napolitano}{1}
\orcidauthor{0009-0004-7069-1326}{B. Moreschini}{2,3,4}
\orcidauthor{0000-0003-2536-1614}{A. Calabr\`{o}}{1}
\orcidauthor{0000-0001-8415-7547}{L. Christensen}{5}
\orcidauthor{0000-0003-1354-4296}{M. Llerena}{1}
\orcidauthor{0000-0002-5268-2221}{T. J. L. C. Bakx}{6}
\orcidauthor{0000-0002-2545-5752}{F. Belfiore}{3,4}
\orcidauthor{0000-0001-8863-2472}{D. Bevacqua}{1}
\orcidauthor{0000-0001-5414-5131}{M. Dickinson}{7}
\orcidauthor{0000-0003-3820-2823}{A. Fontana}{1}
\orcidauthor{0000-0003-3248-5666}{G. Gandolfi}{1}
\orcidauthor{0000-0002-7913-4866}{T. Gasparetto}{1}
\orcidauthor{0000-0002-9889-4238}{A. Marconi}{2,3}
\orcidauthor{0000-0002-9572-7813}{S. Mascia}{8}
\orcidauthor{0000-0001-6870-8900}{E. Merlin}{1}
\orcidauthor{0000-0002-8512-1404}{T. Morishita}{9}
\orcidauthor{0000-0003-2804-0648}{T. Nanayakkara}{10}
\orcidauthor{0000-0002-7409-8114}{D. Paris}{1}
\orcidauthor{0000-0001-8940-6768}{L. Pentericci}{1}
\orcidauthor{0000-0002-0939-9156}{B. P\'{e}rez-D\'{i}az}{1}
\orcidauthor{0000-0002-4140-1367}{G. Roberts-Borsani}{11}
\orcidauthor{0000-0003-2349-9310}{S. Rojas-Ruiz}{12}
\orcidauthor{0000-0002-9334-8705}{P. Santini}{1}
\orcidauthor{0000-0002-8460-0390}{T. Treu}{12}
\orcidauthor{0000-0002-5057-135X}{E. Vanzella}{13}
\orcidauthor{0000-0003-0980-1499}{B. Vulcani}{14}
\orcidauthor{0000-0002-9373-3865}{X. Wang}{15,16,17}
\orcidauthor{0000-0001-9163-0064}{I. Yoon}{18}
\orcidauthor{0000-0002-7051-1100}{J. Zavala}{19}

\affiliation{$^1$INAF Osservatorio Astronomico di Roma, Via Frascati 33, 00078 Monteporzio Catone, Rome, Italy}
\affiliation{$^2$Universit\`{a} di Firenze, Dipartimento di Fisica e Astronomia, via G. Sansone 1, 50019 Sesto Fiorentino, Florence, Italy.}
\affiliation{$^3$INAF – Arcetri Astrophysical Observatory, Largo E. Fermi 5, I-50125, Florence, Italy}
\affiliation{$^4$European Southern Observatory, Karl-Schwarzschild Straße 2, D-85748 Garching bei München, Germany}
\affiliation{$^5$Niels Bohr Institute, University of Copenhagen, Jagtvej 128, DK2200 Copenhagen N, Denmark}
\affiliation{$^6$Department of Space, Earth, \& Environment, Chalmers University of Technology, Chalmersplatsen 4 412 96 Gothenburg, Sweden}
\affiliation{$^7$NSF’s NOIRLab, Tucson, AZ 85719, USA}
\affiliation{$^8$Institute of Science and Technology Austria (ISTA), Am Campus 1, A-3400 Klosterneuburg, Austria}
\affiliation{$^9$IPAC, California Institute of Technology, MC 314-6, 1200 E. California Boulevard, Pasadena, CA 91125, USA}
\affiliation{$^{10}$Centre for Astrophysics and Supercomputing, Swinburne University of Technology, PO Box 218, Hawthorn, VIC 3122, Australia}
\affiliation{$^{11}$Department of Physics \& Astronomy, University College London, London, WC1E 6BT, UK}
\affiliation{$^{12}$Department of Physics and Astronomy, University of California, Los Angeles, 430 Portola Plaza, Los Angeles, CA 90095, USA}
\affiliation{$^{13}$INAF -- OAS, Osservatorio di Astrofisica e Scienza dello Spazio di Bologna, via Gobetti 93/3, I-40129 Bologna, Italy}
\affiliation{$^{14}$INAF -- Osservatorio Astronomico di Padova, Vicolo Osservatorio 5, 35122 Padova, Italy}
\affiliation{$^{15}$School of Astronomy and Space Science, University of Chinese Academy of Sciences (UCAS), Beijing 100049, China}
\affiliation{$^{16}$National Astronomical Observatories, Chinese Academy of Sciences, Beijing 100101, China}
\affiliation{$^{17}$Institute for Frontiers in Astronomy and Astrophysics, Beijing Normal University, Beijing 102206, China}
\affiliation{$^{18}$National Radio Astronomy Observatory, 520 Edgemont Road, Charlottesville, VA 22903, USA}
\affiliation{$^{19}$University of Massachusetts Amherst, 710 North Pleasant Street, Amherst, MA 01003-9305, USA}

\thanks{\vspace{0.1cm}$^\dagger$E-mail: \href{mailto:marco.castellano@inaf.it}{marco.castellano@inaf.it}}

\begin{abstract}
An accurate characterisation of the physical properties of galaxies at cosmic dawn is key to understanding the origin of the high abundance of UV-bright galaxies at z$\gtrsim$10. We exploit deep (9.1-hour exposure time) NIRSpec PRISM observations of GHZ2 to constrain the sources of ionising radiation and the properties of the interstellar medium (ISM) in this bright, compact, and highly ionising galaxy at z=12.3. We measure with high significance the prominent N~IV, C~IV, He~II, O~III, C~III, O~II, and Ne~III emission features previously detected in shallower observations, and confirm the detection of the \niii~multiplet, yielding tight constraints on the N/O ratio, which is found to be $\simeq$2 times the solar value.  
We also detect the \mgii, \feiv~and \sIii~doublets, the \hzetahei~blend, and the \sIivoiv~absorption complex. The \oiiibowen~fluorescence line is only detected in the first observing epoch, implying variability on a rest-frame time span of 19 days, strongly suggesting the presence of an active nucleus. Combining the NIRSpec dataset with available optical and far-infrared constraints from MIRI and ALMA, we show that the emission spectrum of GHZ2 cannot be reproduced by single-density spectro-photometric models, even under extreme assumptions on the ionisation parameter and electron density.
Multi-zone photoionisation modelling performed with the \textsc{HOMERUN} code demonstrates that star formation must be occurring in a strongly stratified ISM, where both low-/intermediate-density gas and high-density regions (log($n_e$/cm$^{-3}) \gtrsim 4$) coexist.
The GHZ2 emission landscape is consistent with either a composite star-formation plus AGN scenario, or with star formation occurring in a combination of radiation- and matter-bounded regions. Purely radiation-bounded stellar models fail to reproduce the observed He~II emission, making an additional hard ionising component unavoidable.
\end{abstract}

\maketitle


\section{Introduction}\label{sec:intro}

The first three years of JWST observations have achieved the discovery \citep[e.g.,][]{Castellano2022b,Naidu2022b,Finkelstein2022b} and spectroscopic confirmation \citep[e.g.,][]{ArrabalHaro2023b,Harikane2024,Napolitano2025} of a high abundance of UV-bright galaxies at z$\gtrsim$10. The spectroscopic investigation of the nature of these distant sources is fundamental to discriminate among competing physical explanations for the \textquote{excess} at the bright-end of the UV luminosity function, in particular concerning a significant contribution to their UV emission from hard-ionising sources such as active galactic nuclei \citep[AGN,][]{Hegde2024,Matteri2025a}, massive or population III stars \citep[e.g.,][]{Kannan2022,Harikane2022b,Trinca2024}. 

The results obtained so far with JWST, mostly using the low-resolution NIRSpec PRISM, have revealed a bimodal high-redshift population composed by a large number of relatively \textquote{ordinary} galaxies with faint/absent UV emission lines, and by objects with prominent emission lines indicative of extreme ionisation conditions \citep[e.g.,][]{ArrabalHaro2023a,Roberts-Borsani2024,Roberts-Borsani2025b,Hayes2025,Tang2025}. The latter category of objects shows high-ionisation features such as \heii, \civ, \ciii, \niv, with equivalent widths (EW) and flux ratios that are consistent either with low-metallicity star formation, or with a composite contribution from both star formation and AGN \citep[e.g.,][]{Maiolino2023,Calabro2024,Napolitano2024b}.  

Several of these sources also show an unexpected super-solar nitrogen abundance, such as GHZ9 \citep[z=10.145,][]{Napolitano2024b}, GN-z11 \citep[z=10.6,][]{Bunker2023},  GHZ2 \citep[z=12.3,][C24 hereafter]{Castellano2024}, and MoM-z14 \citep[z=14.4,][]{Naidu2025}, and they are found to be on average more compact than objects with fainter emission lines \citep{Harikane2025}.

The source of the copious amounts of ionising photons that originate their prominent UV emission spectra is, at present, unknown. A potential scenario is that these objects are  caught during a short, intense star-formation episode including massive, super-massive \citep[e.g.,][]{Denissenkov2014} or very massive \citep[e.g.,][]{Vink2023,Upadhyaya2024,Schaerer2025} stars whose ejecta are responsible for the nitrogen enriched inter-stellar medium (ISM) in their surroundings. In such a scenario, these sources are hypothesized to be precursors of today's globular clusters, whose second generation stars show similar abundance patterns \citep[e.g.,][]{Charbonnel2023,DAntona2023,Senchyna2023,Marques-Chaves2024,Ji2025}. However, clear AGN features have been found in two of these objects, namely GN-z11, which shows the typical AGN lines  \neiv~and CII$^*\lambda $1335 and extreme densities \citep{Maiolino2023}, and GHZ9, which is associated with point-like X-ray emission in Chandra observations \citep{Kovacs2024, Napolitano2024b}. These findings leave open the possibility that a non-negligible contribution to the ionising flux comes from accretion onto a super-massive black hole (SMBH). This scenario is also extremely relevant in the context of the unsolved quest for the initial seeding mechanisms of SMBHs. In fact, the inferred M$_{BH}$ and M$_{BH}$/M$_{star}$ in the aforementioned objects are consistent either with Eddington-limited accretion onto \textit{primordial} black hole seeds, or with super-Eddington accretion on light Population III seeds \citep{Maiolino2023,Huang2024,Dayal2025,Napolitano2024b,Ziparo2025,Matteri2025}.
 
Among these highly-ionising sources, GHZ2 stands out as one of the objects with the best available constraints in terms of available data, benefiting of deep NIRCam, NIRSpec, MIRI, and ALMA observations.  GHZ2 (R.A.=3.4989827 deg., Dec=-30.3247534 deg., $M_{\rm UV} = -20.5$~mag) was first identified by \citet{Castellano2022b} and \citet{Naidu2022b} in multi-band JWST observations of the GLASS-JWST program \citep{Treu2022}, and spectroscopically confirmed by both NIRSpec \citepalias{Castellano2024} and MIRI \citep{Zavala2025} at z=12.3. The analysis of the UV and optical rest-frame dataset shows that GHZ2 has a hard-ionising spectrum (log\,$U\geq$-2), low gas-phase metallicity ($<$15\% solar), a subsolar C/O ratio and a super-solar N/O abundance. However, both UV and optical rest-frame diagnostics fail to discriminate between an AGN and a low-metallicity, high-density, young starburst \citep[][]{Castellano2024,Calabro2024,Zavala2025}. Instead, ALMA follow-up observations point to a star-forming nature of this source \citep{Zavala2024b}. The [O~III]88$\mu$m emission, together with the strong limits on the [O~III]52$\mu$m line, likely originates in a moderate density environment (n$_e<$4000 cm$^{-3}$). Most importantly, the ALMA analysis indicates that GHZ2 follows the same well-established relationships between [O~III]88$\mu$m luminosity and star-formation rate (SFR), and between dynamical mass and H$\beta$ luminosity, as star-forming galaxies and giant H~II regions. 

To summarise, the picture available so far is ambiguous, with GHZ2 showing both typical AGN features, in particular the high EW of the UV lines, and other properties perfectly in line with expectations for low-metallicity star-forming galaxies. Simple nebular line diagnostic diagrams are unable to distinguish between low metallicity star-formation and AGN in GHZ2, and other similar high-redshift objects \citep[e.g.,][]{Ubler2023,Juodzbalis2025}. The adoption of more refined spectro-photometric approaches is thus needed to disentangle AGN and star-formation. The analysis by \citet{ChavezOrtiz2025} based on the BEAGLE-AGN tool \citep{VidalGarcia2024} shows evidence of a significant contribution by an AGN to the UV lines of GHZ2, in particular to the extremely strong C~IV, implying the presence of a log$_{10}$(M$_{BH}$/M$_{\odot}$) $\sim$ 7.20 SMBH at its center. A similar conclusion has been reached by \citet{Zhu2025} that found GHZ2 is consistent with a nitrogen-enhanced AGN on the basis of improved photoionisation models, and by \citet{Fabian2026} that showed that the UV slope of GHZ2 is compatible with emission from a standard accretion disc.

\begin{figure*}[ht]
\centering
\includegraphics[trim={1.8cm 2.0cm 1.8cm 1.5cm},clip,width=18cm,keepaspectratio]{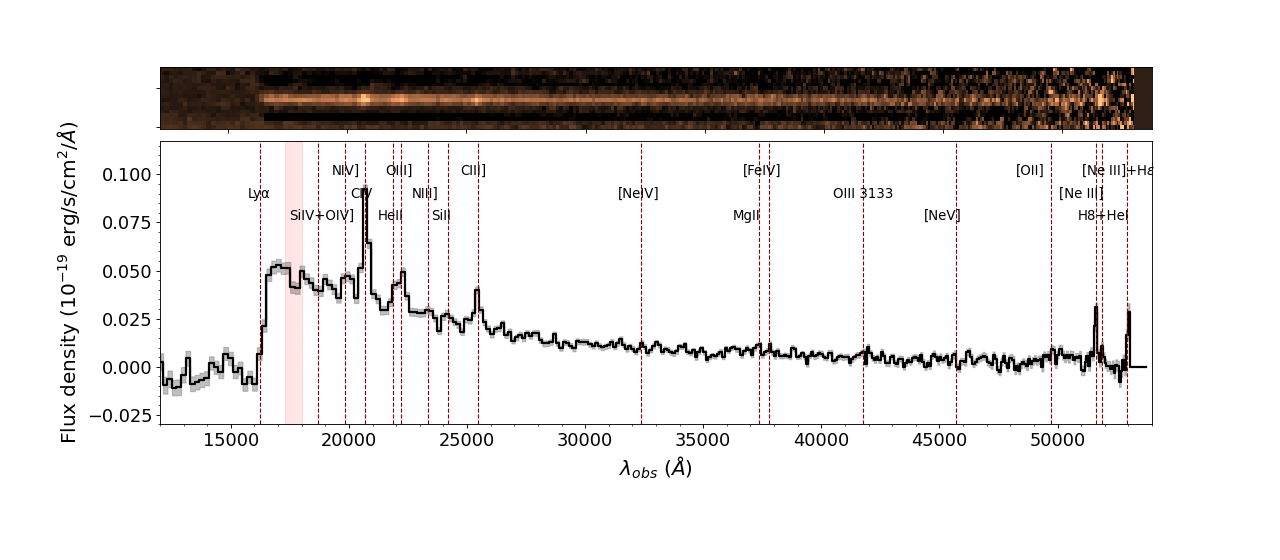}
\caption{Observed 2D (top) and 1D (bottom) NIRSpec PRISM spectra of GHZ2 acquired with a total exposure time of $\sim$9.1 hours. In the bottom panel the gray shaded area shows the 1$\sigma$ uncertainty, the red dashed lines highlight the wavelength of the UV features discussed in the present paper, and the red shaded region encloses the spectral range that has been masked in the present analysis due to background subtraction potentially contaminated by the H$\alpha$ line of a secondary target.} 
\label{fig_FULLSPEC}
\end{figure*} 

In this paper, we will analyse the nature of GHZ2 exploiting the final depth NIRSpec PRISM observations obtained under program GO-3073 (PI M. Castellano), in conjunction with the available MIRI and ALMA observations. The paper is organised as follows: in Sect.~\ref{sec:obs} we describe the data reduction and analysis process; the spectroscopic measurements are discussed in Sect.~\ref{sec:features}, while Sect.~\ref{sec:analysis} presents the results of the spectro-photometric analysis of the source properties under different assumptions on the sources of ionising radiation. We discuss our results in light of the current scenarios for high-ionising, N-enriched sources in Sect.~\ref{sec:discussion}, and present a summary and future prospects in Sect.~\ref{sec:summary}. The public release of the NIRSpec observations, and additional information on the GHZ2 spectrum are described in Appendix \ref{sec:appendix-datarelease} and Appendix \ref{sec:appendix-addinfo}, respectively.

Throughout the paper we adopt AB magnitudes \citep{Oke1983}, a solar metallicity of 12 + log(O/H) = 8.69 \citep{Asplund2009}, and a flat $\Lambda$CDM concordance model (H$_0$ = 70.0~km~s$^{-1}$ Mpc$^{-1}$, $\Omega_M=0.30$).

\section{Observations and data analysis}\label{sec:obs}
A detailed description of the GO-3073 NIRSpec PRISM observations, data reduction and public data release is presented in Appendix~\ref{sec:appendix-datarelease}. Here we provide a summary of the elements that are most relevant for the analysis of GHZ2. 

In the present paper we exploit NIRSpec observations of GHZ2 acquired in two different epochs for a total observing time of 32825~s, which were reduced as described in \citetalias{Castellano2024} \citep[see also][]{Napolitano2025,Napolitano2024b}.  Two out of three nodding positions of the third visit in the first epoch (i.e., \textquote{P12}) were affected by an electric short and are unusable. We decided to exclude P12 from the final stack considering the negligible contribution in terms of exposure time and the increased uncertainty in the background subtraction on a single dither position \citepalias[see][]{Castellano2024}. We also produced a stacked spectrum of the second epoch visits (acquired on 2024 July 3-4, total observing time of 15323~s) to be compared to the first epoch observations acquired on 2023 Oct. 24 and discussed in \citetalias{Castellano2024}. Before performing a standard three-nod pattern background subtraction we masked spectral regions contaminated by a secondary target at $z_{spec}$=1.68 which is visible in the upper part of the slit in one of the dithered positions of the first epoch visits \citep[ID=39116 at R.A.=3.4990961 deg., Dec=-30.3249619 deg in][]{Merlin2024}. Nonetheless, we will not consider in the analysis the region of the H$\alpha$ line of the secondary object that may affect the GHZ2 spectrum at 1310--1330~\AA\ rest-frame. The final spectrum is shown in Fig.~\ref{fig_FULLSPEC}. 

For consistency with the \citetalias{Castellano2024} analysis, we use a linear relation of the ratio between NIRCam photometry and synthetic NIRSpec photometry to correct for wavelength-dependent slit and aperture losses. We verified that line measurements do not significantly change when adopting an average correction \citep[e.g.,][]{Napolitano2025} in place of a wavelength-dependent one \citep[see also][]{Roberts-Borsani2024, Napolitano2025c}. Throughout the work we will derive rest-frame physical properties taking into account that GHZ2 is affected by moderate lensing magnification ($\mu = 1.3$) estimated on the basis of the model by \citet{Bergamini2023} \citep[see also the discussion about the lens modelling in][]{Zavala2025}. 
 
We first computed the spectroscopic redshift of GHZ2 on the final spectrum from a weighted average of the centroids of the best resolved, high-signal-to-noise ratio (SNR) lines (\niv, \civ, \ciii, and \neiii), obtaining $z=12.341 \pm 0.005$, in perfect agreement with the estimate by \citetalias{Castellano2024}. We then visually inspected the spectrum to assess consistency with the detected lines presented by C24 and search for other potential spectral features that may have been undetected in the shallower first-epoch observations. We assessed the significance of the detected lines, and measured their fluxes and EW as described by \citet{Napolitano2025}. Briefly, we consider significant all the features with SNR $\geq$3 as evaluated from direct integration of the continuum-subtracted spectrum in a window centred at the expected wavelength $\lambda$ and having a width $4\times \sigma_R (\lambda)$, where $\sigma_R (\lambda)$ is the expected Gaussian root-mean-square (RMS) of a line observed at resolution $R(\lambda$)\footnote{The instrumental resolution is obtained from \url{https://jwst-docs.stsci.edu/jwst-near-infrared-spectrograph/nirspec-instrumentation/nirspec-dispersers-and-filters} with the assumption of a source that illuminates the slit uniformly.}. In the case of partially blended lines, we assess the significance of the entire line complex and of the different components in windows with width $2 \times \sigma_R (\lambda)$. The continuum was estimated through a linear interpolation of regions closest to each line that are free from potential features, using the \textsc{emcee} \citep[][]{Foreman_Mackey2013} routine. For all significant emission features, we performed a Gaussian fit on the continuum-subtracted flux using the \textsc{specutils} package of \textsc{astropy} \citep{Astropy2013}. The uncertainty in the continuum is taken into account in the Gaussian fit. Unresolved doublets and multiplets were fitted as a single Gaussian profile, while partially blended lines were fitted with a double-Gaussian profile. The centroid of the Gaussian was allowed to vary with $\Delta$z = 0.04, and the Gaussian standard deviation within 5\% of the nominal $\sigma_R(\lambda)$, to account for redshift and calibration uncertainties, and for the presence of unresolved multiplets. The best model parameters and the integrated flux were determined by taking the median of the posterior distributions resulting from a Markov chain Monte Carlo (MCMC) analysis performed with \textsc{emcee}. Uncertainties were calculated based on the 68-th percentile posterior density intervals.

\section{Spectroscopic features in the GHZ2 spectrum}\label{sec:features}

We measure at high significance the UV emission features detected in the first-epoch observations. In particular the prominent \civd~(EW$\sim$36~\AA) and \ciiid~(EW$\sim$17~\AA) doublets, and the high-ionization \heii~(EW$\sim$8~\AA) and \nivd~(EW$\sim$9~\AA) lines. Most importantly, we confirm the detection of the \niii~multiplet and place stringent 3$\sigma$ upper limits of $\sim 10^{-19}$ \ergs\ on the flux, and $\sim$10 \AA~on the EW, for the very high-ionization \neiv~and \nev\ lines. The Ly$\alpha$ is not detected, with a 3$\sigma$ limit of EW(Ly$\alpha$) $ < 6.3$~\AA~estimated as in \citetalias{Castellano2024} with the procedure described in detail by \cite{Jones2023C} and \cite{Napolitano2024}. All the line fluxes and EW measured on the full-depth spectrum are in agreement within 2$\sigma$ with the first-epoch measurements, with the exception of \nivd, and \civd, for which we measure consistent fluxes but $\approx$20\% lower EWs, and of \neiii, which is found to be a factor of $\sim1.5$ brighter. 
\begin{table}[ht]
\caption{UV emission lines in GHZ2}\label{tab:lines}
\centering
\begin{tabular}{ccc}
\hline
Line& Flux & EW \\
 & ($10^{-19}$~erg~s$^{-1}$~cm$^{-2}$) & (\AA) \\
 \hline
\sIivoiv & - & -3.5 $\pm$  1.2\\
\niv & 6.7 $\pm$  0.4 & 9.3  $\pm$  0.6\\
\civ & 26.2   $\pm$   0.4  & 36.1   $\pm$   1.1\\
\heii & 5.2   $\pm$   0.4  & 8.4  $\pm$    0.6 \\
\oiii & 8.8  $\pm$    0.4  & 14.4  $\pm$    0.7 \\
\niii & 3.2  $\pm$    0.3  & 6.1  $\pm$    0.6  \\
\sIii & 3.0  $\pm$    0.3  & 6.4  $\pm$    0.7  \\
\ciii & 8.2  $\pm$    0.3  & 17.1  $\pm$    1.0 \\
\neiv & $<$0.42 & $<$1.9 \\
\mgii &  2.6   $\pm$ 0.2 & 14.2 $\pm$ 1.2\\
\feiv &  1.6  $\pm$ 0.2 &9.3  $\pm$ 1.1\\
\oiiibowen~(first epoch) & 2.9  $\pm$    0.5  & 30.5  $\pm$    10.3 \\
\oiiibowen~(second epoch) & $<$0.9  & $<$9 \\
\nev &  $<$1.0  & $<$9\\
\oii & 1.8  $\pm$   0.3  & 14  $\pm$   3  \\
\neiii & 9.3   $\pm$   0.4  & 75   $\pm$   7 \\
\hzetahei &  1.3   $\pm$ 0.3 & 10 $\pm$ 2\\
\hline
\end{tabular}
\end{table}
We attribute these differences to the improved estimate of continuum emission, in particular in the case of \neiii~which falls at the edge of the first-epoch spectrum, although we cannot exclude that the different orientation and centring of the second-epoch observations (Fig.~\ref{fig_GHZ2appendix}) may slightly affect the ratio between line and continuum flux in the extracted spectrum. 

We also measure a significantly lower flux and EW for the \oiiibowen~fluorescence line, which we discuss in more detail in Sect.~\ref{subsec:Bowen}. 

The final spectrum yields the detection of \sIiid, \mgiid~and \feivd~doublets, and of the \hzetahei~blend, which were not discussed in \citetalias{Castellano2024}. The [Ne~III]$\lambda$3967+H$\epsilon$ blend is also well detected, but it is truncated at the edge of the extracted spectrum, which prevents a measurement of its flux. The \hzetahei~blend was at the very edge of the \citetalias{Castellano2024} spectrum while [Ne~III]$\lambda$3967+H$\epsilon$ was not covered by previous observations. We verified that the remaining features were significant at the $\sim$2-4$\sigma$ level in the first epoch dataset, with fluxes consistent within uncertainties with the ones reported here.

The inspection of the region blueward of N~IV] also reveals the presence of the \sIivoiv~absorption feature which may be of ISM, stellar wind or photospheric origin. The feature is found to be significant at the $>$3$\sigma$ level by direct integration of the continuum-subtracted spectrum, and has a EW=-3.5 $\pm$ 1.1 \AA, consistent with measurements in deep spectra of z$\sim$2 Lyman-break galaxies \citep[e.g.,][]{Talia2012}. In Table~\ref{tab:lines} we report the fluxes and rest-frame EWs measured from the Gaussian fits (Sect.~\ref{sec:obs}) of all the detected features. Detailed views of individual emission lines are shown in the appendix (Fig.~\ref{fig_lines}).

\subsection{The Ly$\alpha$ damping wing}\label{subsec:dampingwing}
The high SNR ($\sim$15) at the position of the Ly$\alpha$ break enables the analysis of the damping wing profile of GHZ2. The emission of the source around the rest-frame Ly$\alpha$ damping wing is influenced by the nature of its UV emission, whether it originates from hot, young stars, the nebular continuum, or AGN. In addition, the intrinsic spectral slope may be reddened by dust attenuation. Both have a strong influence on the estimate of the intrinsic continuum emission, and therefore on the overall shape of the Ly$\alpha$ damping wing.

We measure the slope of the continuum emission blueward of the rest-frame wavelength 1500~{\AA} from the best-fit \textsc{BAGPIPES} model described in Sect.~\ref{sec:bagpipes}. The \textsc{BAGPIPES} model includes a reddening of $A_V = 0.29\pm0.14$ with a \citet{Cardelli1989} Milky Way extinction curve, so to recover the intrinsic emission we correct the extrapolated power-law spectral slope for the reddening, and use this model power-law function to normalise the observed spectrum of GHZ2. We then fit the Ly$\alpha$ line profile of the normalised spectrum between 1200~{\AA} and 1500~{\AA} rest-frame with a model that contains both a contribution of absorption from the intergalactic medium with a range of neutral gas fraction, $x_{HI}$ between 0 and 1 \citep{Miralda1998} and an intrinsic HI absorption component close to the source, convolved at the PRISM spectral resolution of $R\sim35$ around 1.7$\mu$m.
Both components are fixed to the redshift of GHZ2. We find  a neutral gas component with a column density of log~N(HI)/cm$^{-2}$= 22.35$\pm$0.37, while the neutral hydrogen fraction in the IGM is found to be unconstrained ($x_{HI}=0.63_{-0.63}^{+0.37}$). The lack of constraining power on $x_{HI}$ is not surprising, considering the strong degeneracy between the two parameters \citep{Huberty2025}. It is reassuring that consistent results for N(HI) are obtained when fixing $x_{HI}=1$ in the fit, which is a reasonable assumption at z=12.3 \citep[e.g.,][]{Nakane2024,Umeda2025}. In addition, consistent results are obtained when using an SMC extinction curve in the \textsc{BAGPIPES} fit, demonstrating little dependence of the derived N(HI) on the details of the UV continuum modeling.

The column density measured for GHZ2 is comparable to the one measured for JADES-GS-z14-0 \citep[][]{Heintz2025b}, and in line with recent findings of a high fraction of sources with strong integrated DLA at z$>$8 \citep{Heintz2025}, including objects with log~N(HI)/cm$^{-2}$$>$22 (\citealt{Heintz2024}, but see also \citealt{Mason2025}). Simulations predict that these strong galaxy-DLAs are associated with the dense ISM close to the
star-forming regions in M$_h\gtrsim$10$^{11}$M$_{\odot}$ halos \citep[][]{Gelli2025}, with sightline-to-sightline column density variations on scales comparable to the size of GHZ2 \citep[$\lesssim$100 pc,][]{Yang2022b,Ono2023}. The large N(HI) in front of GHZ2 is consistent with the stringent non-detection of the Ly$\alpha$ line (Sect.~\ref{sec:features}), and it is suggestive of a large gas reservoir feeding its massive and compact star-formation episode in a scenario of rapid galaxy assembly  \citepalias{Castellano2024}.

\subsection{Evidence of variability of the Bowen fluorescence emission at 3133~\AA}\label{subsec:Bowen}
\begin{figure}
\centering
\includegraphics[trim={1.5cm 2.0cm 0.55cm 1.4cm},clip,width=\linewidth,keepaspectratio]{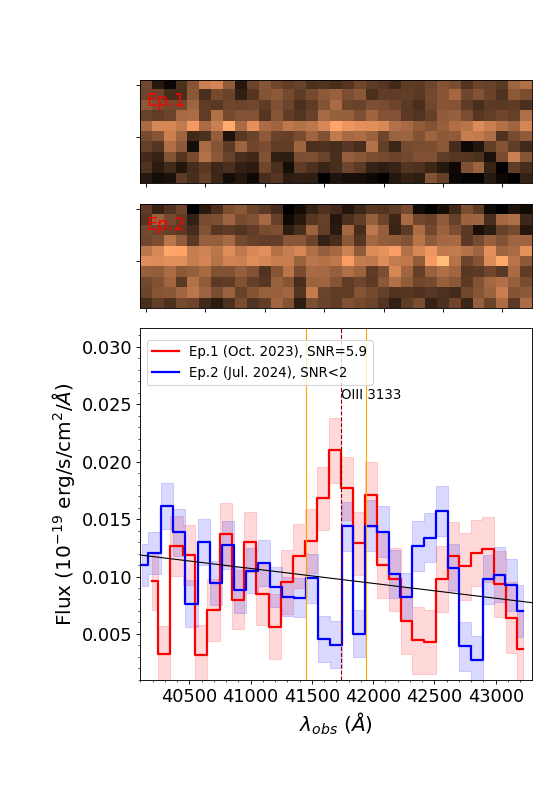}
\caption{NIRSpec 2D (top) and 1D (bottom) spectra in a region with width of 240 \AA~rest-frame centred at the position of the \oiiibowen~line. The red and blue lines and shaded areas in the bottom panel show respectively the first and second epoch spectra and relevant 1$\sigma$ uncertainties in each pixel. The vertical orange lines enclose the region where the signal-to-noise ratio (SNR) of the feature is evaluated from direct integration. The black line shows the UV continuum estimated on the stacked spectrum. The red dashed line marks the wavelength of the 3133\AA~ \ion{O}{3} Bowen fluorescence feature. The corresponding regions of the 2D spectra are shown on top.} 
\label{fig_OIIIBowen}
\end{figure}

The first-epoch dataset of \citetalias{Castellano2024} presented a significant detection of the O~III line at 3133~\AA\ emitted via Bowen resonance fluorescence \citep{Bowen1934,Bowen1947}. This line is generated in a highly ionized and dense environment by the resonant decay of He~II Ly$\alpha$ photons and is commonly observed in Seyfert nuclei \citep[e.g.,][]{Netzer1985,Malkan1986,Schachter1990}, symbiotic and X-ray binaries \citep[e.g.,][]{Schachter1989,Selvelli2007} and planetary nebulae \citep[e.g.,][]{Liu1993}.  We found that in the stacked, full-depth spectrum the emission is detected with a lower significance and flux (1.6$\pm$0.3 $10^{-19}$~\ergs), compared to first epoch observations \citepalias[2.9$\pm$0.5 $10^{-19}$~\ergs][]{Castellano2024}. We thus compared the spectra of the two separated epochs acquired in Oct. 2023 and July 2024, respectively, finding that the lower SNR and flux in the stacked spectrum is due to the emission feature being undetected in the latter, with a 3$\sigma$ upper limit of 0.9 $10^{-19}$~\ergs. The discrepancy between the two epochs cannot be explained by a simple measurement scatter. In fact, by randomly extracting fluxes according to the first epoch flux and uncertainty, we estimated a probability of 3$\times 10^{-5}$  of measuring a value equal, or lower than the second epoch upper limit. The comparison between the two epochs at the position of the \oiiibowen~line is shown in~Fig.~\ref{fig_OIIIBowen}. 

As a first check, we visually inspected the individual 2D nods generated by the Step2 pipeline to verify that no residual artefacts were present at the position of the line. We then analysed an independent reduction of both the final spectrum and of the separated epochs to ensure that the difference between the two datasets is not caused by erroneous handling of the reduction process. This alternative data reduction has been performed through the procedure described in detail by \citet{Roberts-Borsani2024} \citep[see also,][]{Roberts-Borsani2025b} which is based on both official STScI pipeline routines and custom codes. We found the same results as in our reference spectra, with the line detected at SNR$>$5 in the first epoch dataset and undetected in the second epoch dataset, resulting in a SNR$\sim$3 detection in the stacked spectrum (left panel in Fig.~\ref{fig_OIIIBowen_visitsGRB}). Finally, we inspected reduced spectra of the five separated pointings comprising our dataset, finding a detection on both P10 and P11 observations of Oct. 2023 at a SNR consistent with expectations, and no detection in any of the July 2024 pointings (Fig.~\ref{fig_OIIIBowen_visitsGRB}, right panel). We thus conclude that the non-detection of this feature in the second-epoch observations is most likely  attributed to intrinsic variability of the emission line. 

The variability can only be due to the emission line originating from a small spatial region of GHZ2, as the time elapsed between the two epochs is only $\sim$19 days rest-frame. In turn, this most likely implies the presence of an AGN, similarly to the recently observed class of flaring super-massive black holes \citep[\textquote{Bowen fluorescence flares}][]{Trakhtenbrot2019,Makrygianni2023} that show short time-scale variability of the \oiiibowen~emission. 
We have checked whether variability is also apparent in the photometric observations of GHZ2. To this aim, we analysed the NIRCam LW observations of the GLASS-JWST field \citep{TreuGlass22} acquired in three different epochs, namely on 2022 June 29; 2022 Nov. 11 and on 2023 July 7. We reduced the F277W, F356W, and F444W observations as described in \citet{Paris2023} \citep[see also][]{Merlin2022,Merlin2024} using the latest calibration files available at the time of writing. We used \textsc{A-PHOT} \citep{Merlin2019b} to measure fluxes of all sources in the field in 0.2 arcsec diameter apertures. While we find GHZ2 to be $\sim$5-10\% brighter in the first two epochs than in the third, the offset cannot be considered as significant. In fact, it is consistent within $\sim$2$\sigma$ with a flux ratio of 1, and falls within the observed photometric scatter computed from other GLASS-JWST sources of similar aperture magnitude. We thus conclude that any continuum variability on the observed time span is at most at the $\sim$10\% level. Thus, we consider the broad band photometry inconclusive. 

\subsection{Line diagnostic diagrams}\label{subsec:diagnostics}
We used the updated line measurements to inspect the position of GHZ2 in diagnostic diagrams exploiting ratios and EWs of UV lines to discriminate among star-formation, AGN, and shocks as main emission mechanism, and compared it to predictions from the emission models by \citet[][]{Gutkin2016}, \citet[][]{Feltre2016} and \citet[][]{Nakajima2022}. Consistently with the \citetalias{Castellano2024} analysis, we find that the nature of the ionising source in GHZ2 cannot be firmly assessed through standard diagnostic diagrams. We show all considered diagrams in the Appendix (Fig.~\ref{fig_DIAGRAMS}). Briefly, in most of the cases, the object falls in regions compatible with both low-metallicity star-formation and AGN emission, e.g. in the widely used C~III]/He~II versus O~III]/He~II, C~IV/C~III] versus (C~IV+C~III])/He~II, and C~III]/He~II versus O~III]/He~II flux ratio diagnostics \citep[e.g.,][]{Feltre2016,Mingozzi2023}. The rest-frame EWs of the UV lines are indicative of AGN or AGN plus star-formation composite emission in the EW(C~III]) versus C~III]/He~II, the EW(C~IV) versus C~IV/He~II, the EW(O~III]) versus O~III]/He~II diagrams according to the criteria by \citet{Nakajima2018a} and by \citet{Hirschmann2019}. The GHZ2 line ratios are also compatible with emission by shocks according to the criteria by \citet{Mingozzi2023} (C~III]/He~II vs. O~III]/He~II), consistently with the analysis by \citet{Flury2024} based on the modeling of N~IV]/C~III] and N~III]/C~III] versus O~III]/He~II.

\section{An in-depth analysis of the source of ionising photons and ISM properties in GHZ2}\label{sec:analysis}

The unusual UV spectrum of GHZ2 is suggestive of extreme ISM conditions and/or of an AGN contribution, which, in turn, may also explain the short-term variation of the Bowen fluorescence emission. To progress beyond the usual diagnostic diagrams, which provide a limited and uncertain assessment on the nature of the ionising conditions, we describe in the following an in-depth analysis aimed at disentangling the potential contribution of emission from a high-density ISM or accretion.

\begin{figure}
\centering
\includegraphics[trim={0.3cm 9.0cm 0.2cm 0.2cm},clip,width=\linewidth,keepaspectratio]{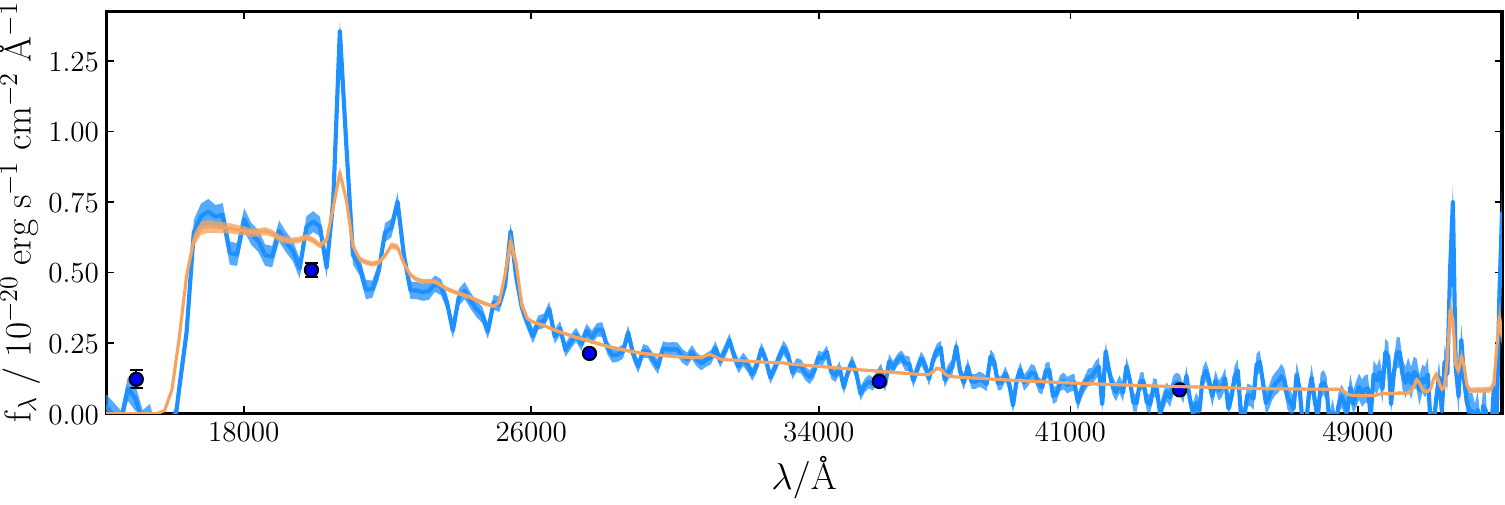}
\includegraphics[trim={0.3cm 0.3cm 0.2cm 0.2cm},clip,width=\linewidth,keepaspectratio]{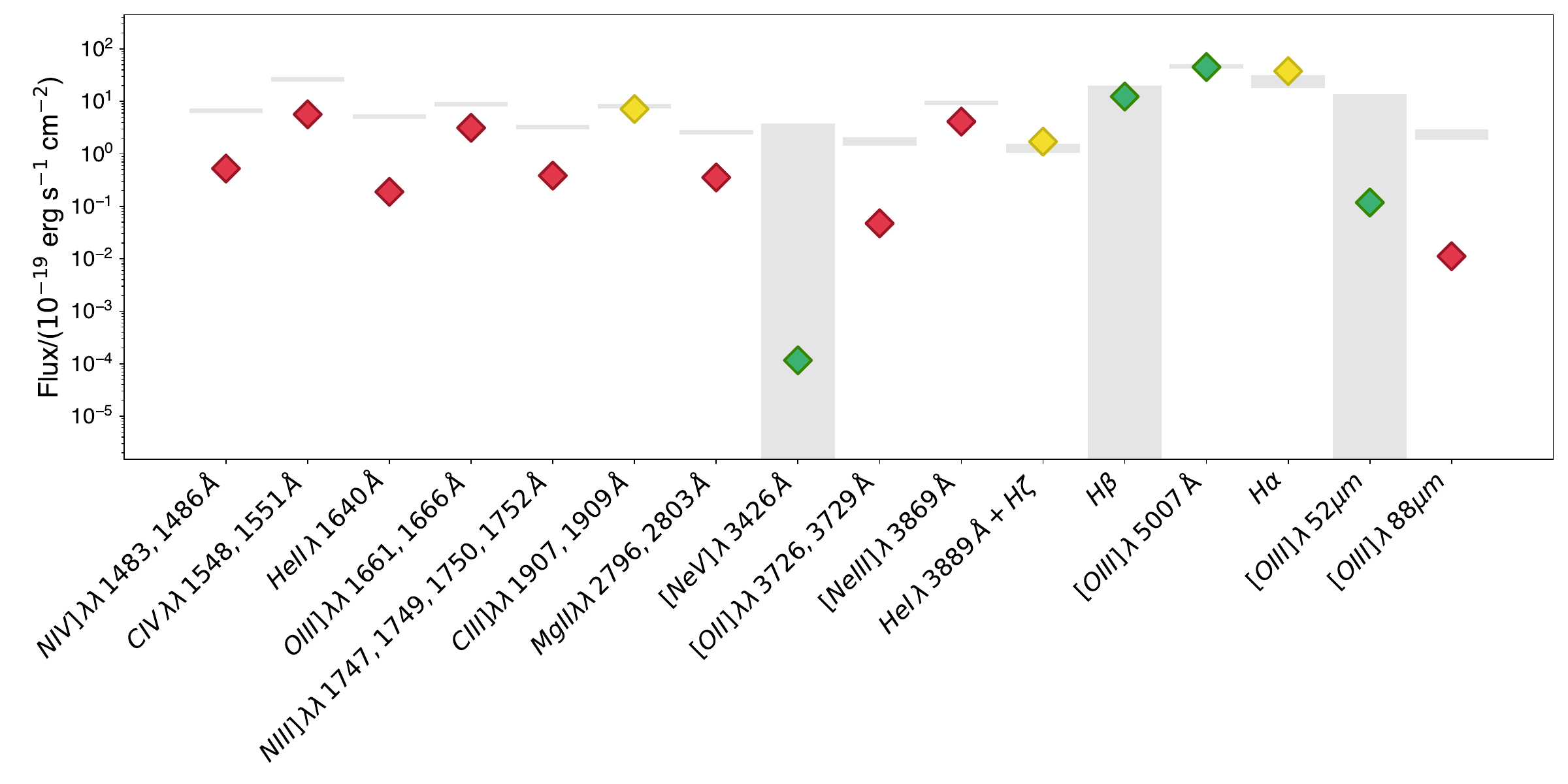}
\caption{\textbf{Top:} Best-fit \textsc{BAGPIPES} template (yellow) obtained by performing a spectro-photometric fitting on the observed NIRSpec spectrum (blue) and NIRCam photometry (dark blue circles and errorbars) of GHZ2, using BPASS v. 2.2.1 stellar models, nebular emission computed with  \textsc{CLOUDY} by assuming electron density log(n$_e$)=5, and a double power-law SFH. \textbf{Bottom:} comparison between the predicted line fluxes obtained by \textsc{BAGPIPES} (diamonds) and the observed ones (grey shaded areas). A minimum uncertainty of 10\% is considered for the observed fluxes. Predicted fluxes are shown as green, yellow, orange, or red diamonds if they are at $\leq$1$\sigma$, $\leq$2$\sigma$, $\leq$3$\sigma$ or $>$3$\sigma$ from observed ones.} 
\label{fig_bagpipes_fit}
\end{figure} 

\subsection{Spectro-photometric fitting with BAGPIPES}\label{sec:bagpipes}

The analysis by \citet{ChavezOrtiz2025} has shown that a simple spectro-photometric fitting with \textsc{BAGPIPES} is unable to reproduce the spectrum of GHZ2 assuming cloud densities of 10$^{2}$ cm$^{-3}$ and 10$^4$ cm$^{-3}$, suggesting that an AGN component is required. Here we extend their tests considering template libraries with associated nebular emission computed by assuming log(n$_e$/cm$^{-3}$)=3, and log(n$_e$/cm$^{-3}$)=5, and allowing the ionisation parameter to cover the range $-3 \leq$log\,$U < 1$. Our goal is to provide a quantitative comparison to the multi-zone analysis presented in Sect.~\ref{sec:specfit}, by finding the \textsc{BAGPIPES} model parameters that best reproduce the GHZ2 emission spectrum, including the UV features presented in Sect.~\ref{sec:features}, the MIRI measurements of \oiiioptred, H$\beta$ and H$\alpha$ reported by \citet{Zavala2025} and the ALMA observations of the [O~III]52$\mu$m and [O~III]88$\mu$m lines by \citet{Zavala2024b}.
We use \textsc{BAGPIPES} v. 1.0.3 \citep{Carnall2018,Carnall2019b} with templates based on \textsc{BPASS} v. 2.2.1 stellar models with an upper-mass cutoff of the IMF of 300~M$_{\odot}$ \citep{Eldridge2017,Stanway2018}. The nebular emission is computed self-consistently with \textsc{CLOUDY} \citep{Ferland2013} as described by \citet{Carnall2018}. We constrained the gas metallicity in the range 0.04-0.15 \zzsun~ indicated by NIRSpec and MIRI analysis \citep{Calabro2024}. We tested two different assumptions on the star-formation history, namely a double power-law model and the non-parametric star-formation history by \citet{Iyer2019} with five lookback time bins, and two different dust attenuation laws, namely \citet{Calzetti2000} and \citet{Cardelli1989}. 

We find that the fit that best reproduces the observed emission lines is obtained with the double power-law SFH, \citet{Cardelli1989} attenuation, and an electron density log(n$_e$/cm$^{-3}$)=5. We show in Fig.~\ref{fig_bagpipes_fit} the best-fit \textsc{BAGPIPES} model (top panel) and the comparison between the observed and modeled line fluxes (bottom panel). 

We find that \textsc{BAGPIPES} underpredicts the observed line fluxes, despite the fit being clearly oriented at maximising the ionising output in terms of age, metallicity and ionisation parameter, in agreement with the analysis by \citet{ChavezOrtiz2025}. The best-fit star-formation history indicates an extremely young object with only a rising SFH component of age$\simeq$4.5 Myr, a metallicity of 0.05$\pm$0.01 \zzsun, and an extreme log\,$U \simeq 0.6$.  We find that the fit has \oiiioptred, H$\beta$ and \hzetahei~fluxes within the observed 1$\sigma$ uncertainty, and predicts \neiv~and [O~III]52$\mu$m emission consistent with the relevant upper limits. The \ciii~and H$\alpha$ lines are marginally consistent within 2$\sigma$ with the measured values. However, \textsc{BAGPIPES} underestimates the [O~III]88$\mu$m flux and all the remaining UV emission lines. Most notably, the \civ~and \oiii~total emission is underpredicted by a factor of $\sim$4-5, the \niv, \heii~ and \oii~emission by a factor of $\sim$10, $\sim$30, and $\sim$40, respectively. The best-fit model obtained with the \citet{Iyer2019} SFH and log(n$_e$/cm$^{-3}$)=5, which is characterized by an ongoing burst of SFR$\sim$10 M$_{\odot}$/yr and an older stellar component of age$\sim$100 Myr, provides a poorer fit to the observed line fluxes, with the \ciii~emission also underestimated at the $>$3$\sigma$ level.  Similarly, all the fits performed under the assumption of log(n$_e$/cm$^{-3}$)=3 fail to match any of the observed lines at $>$3$\sigma$ level. We verified that the results do not change when leaving metallicity free to vary. Unsurprisingly, the fits provide significantly different ranges for the physical parameters, with the stellar mass found in the range log\,$(M_{\rm star}/{\rm M}_{\odot}) = 8.4$ (double power-law SFH) to $\simeq$9.3 (\citealt[][SFH]{Iyer2019}), and the sSFR from $\sim$1 to $\sim$10 Gyr$^{-1}$. While these values are all consistent with the range discussed by \citetalias{Castellano2024} and \citet{Zavala2025}, they should be taken with caution considering the very poor fit of the observed emission spectrum discussed above. 

Most importantly, these tests show that even under the most extreme assumptions in terms of stellar population templates, age, ISM density and ionization parameter, a simple stellar plus nebular model is unable to match the prominent UV emission features and, at the same time, the observed fluxes of optical and FIR lines of GHZ2.

\begin{figure}
\centering
\includegraphics[trim={0.cm 2.0cm 0.0cm 3.5cm},clip,width=\linewidth,keepaspectratio]{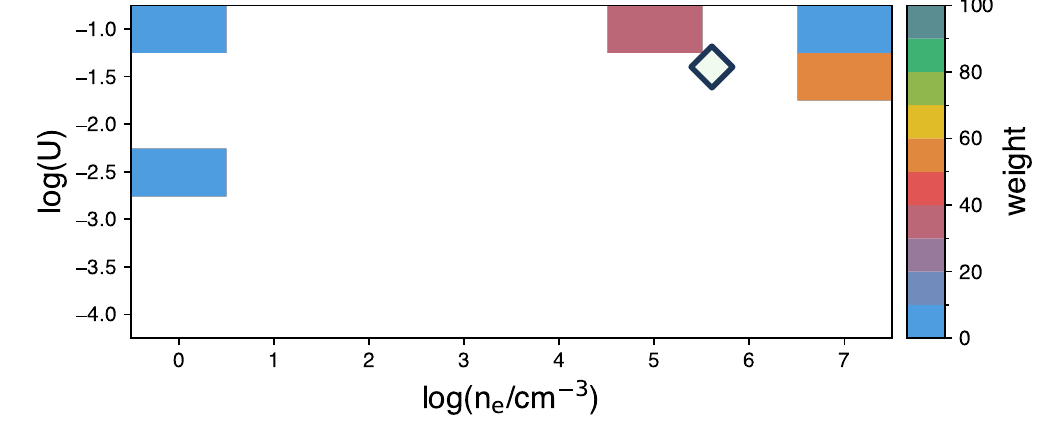}
\includegraphics[trim={0.0cm 0.0cm 0.0cm 0.0cm},clip,width=\linewidth,keepaspectratio]{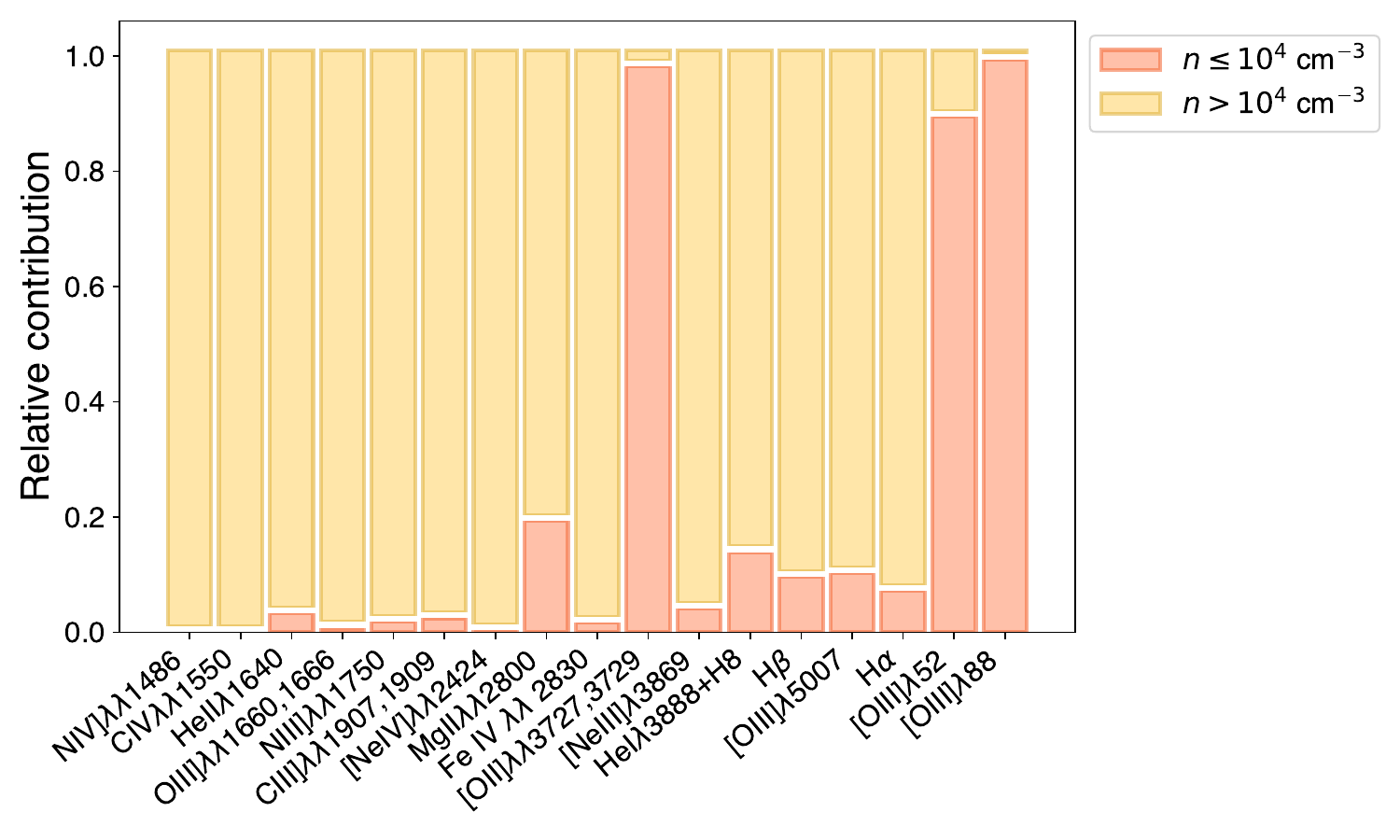}
\includegraphics[trim={0.0cm 0.0cm 0.0cm 0.0cm},clip,width=\linewidth,keepaspectratio]{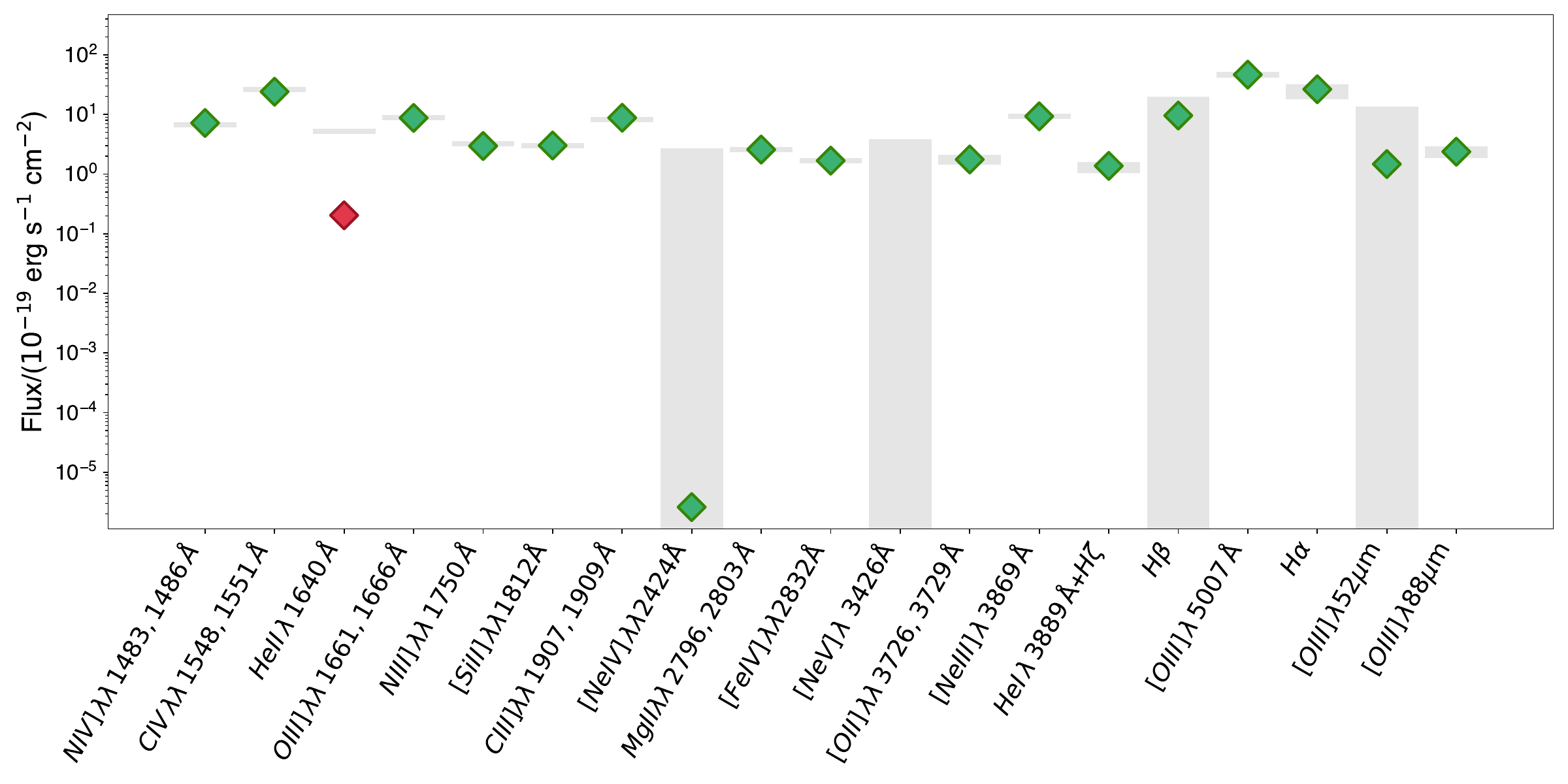}
\caption{Results for the \textsc{HOMERUN} M1 model of GHZ2 (radiation-bounded star-formation case). \textbf{Top panel:} the best-fit grid of single-cloud models in the log\,$U$ versus log(n$_e$/cm$^{-3}$) plane. The cells are colour-coded according to the assigned weight (when this is $>$0) of the relevant single-cloud model, as indicated by the colour bar. The black empty diamond represents the weighted density and ionization parameter of the single-cloud models. \textbf{Central panel:}  relative contribution from high-density ($n>10^4\,\mathrm{cm}^{-3}$, yellow) and low-density (red) star-forming regions. \textbf{Bottom panel:} comparison between the predicted (diamonds) and observed (grey shaded areas) line fluxes (colour labels as in Fig.~\ref{fig_bagpipes_fit}). }\label{fig_homerun_SF}
\end{figure}

\begin{figure}
\centering
\includegraphics[trim={0.cm 0.0cm 0.0cm 0.0cm},clip,width=\linewidth,keepaspectratio]{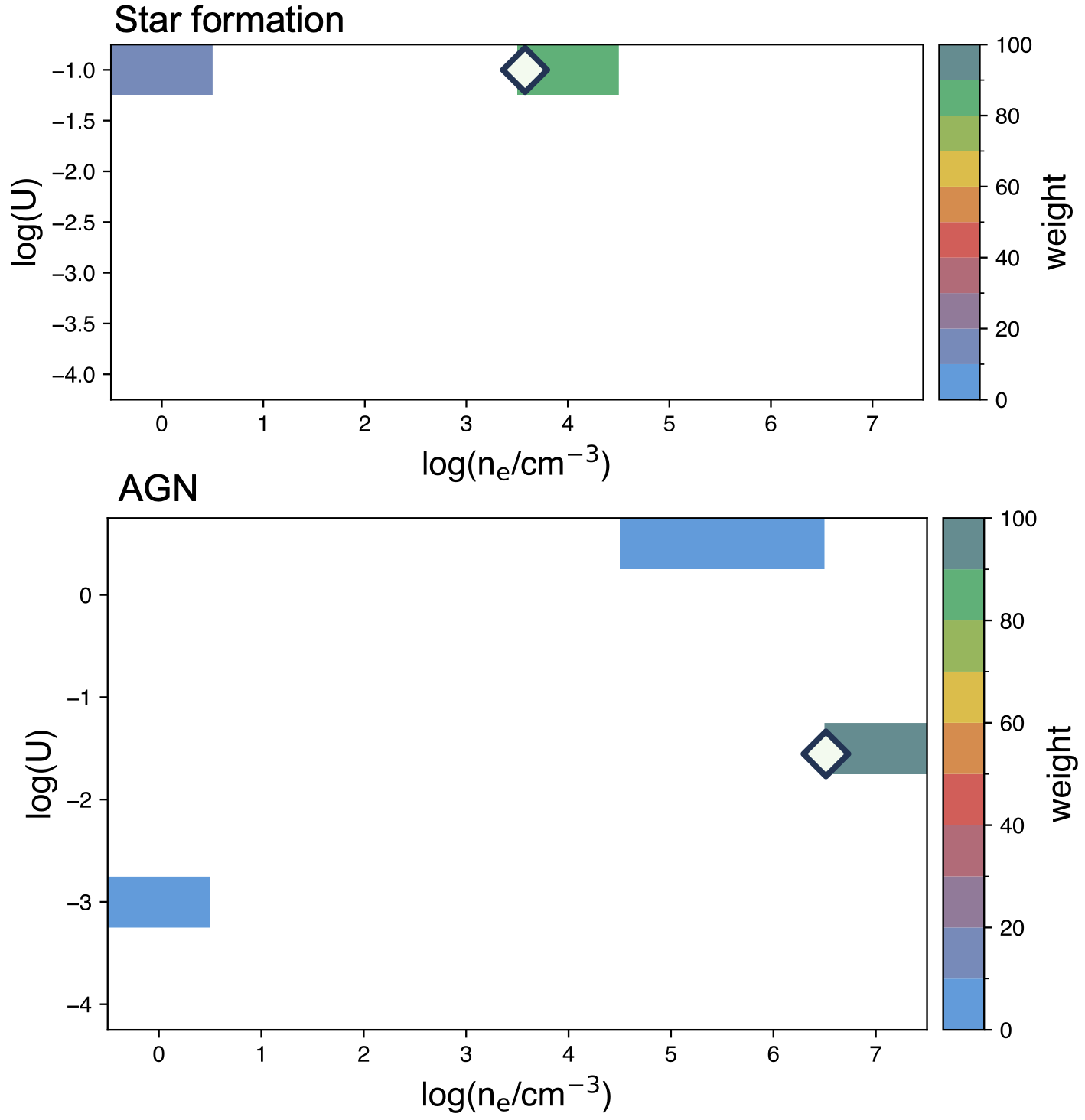}
\includegraphics[trim={0.0cm 0.0cm 0.0cm 0.0cm},clip,width=\linewidth,keepaspectratio]{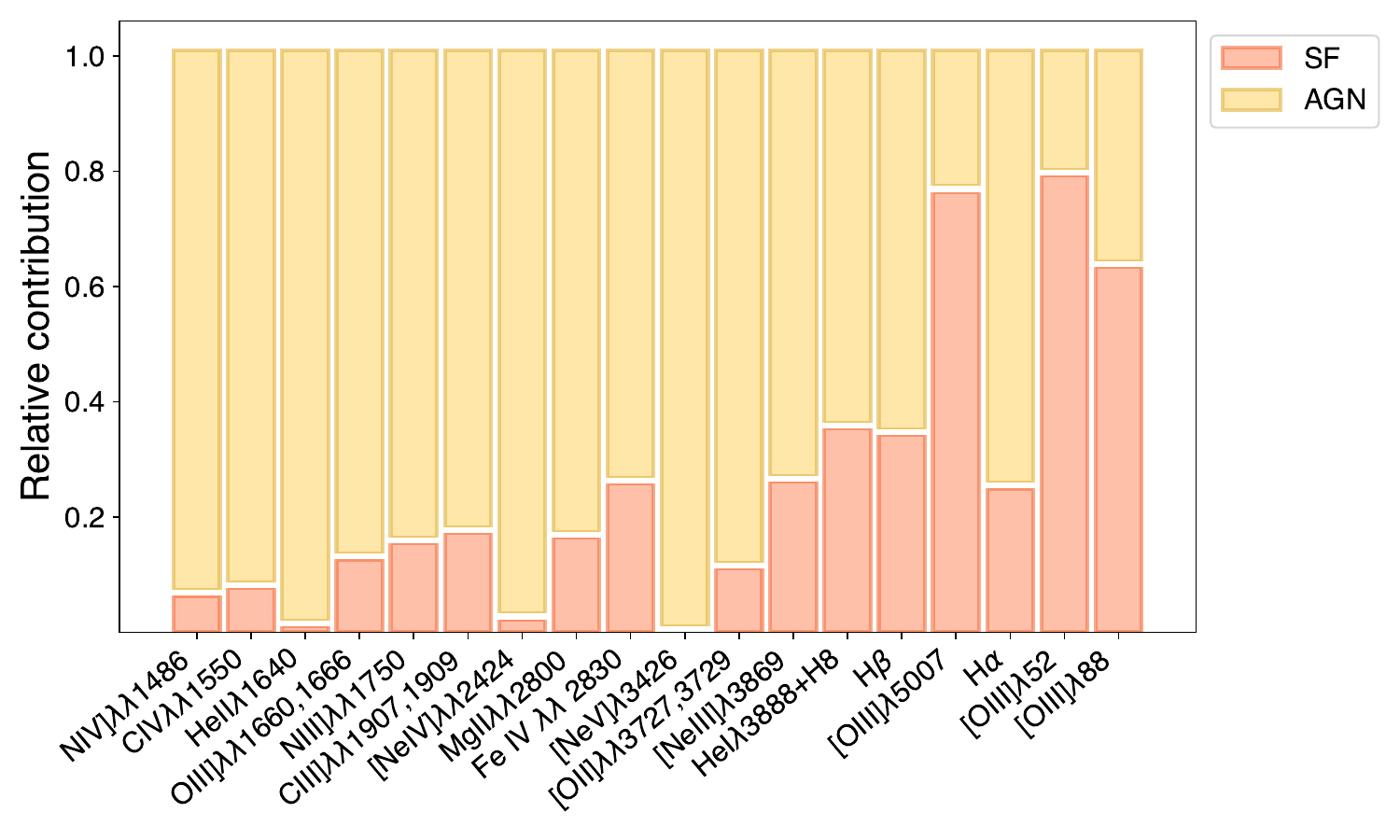}
\includegraphics[trim={0.0cm 0.0cm 0.0cm 0.0cm},clip,width=\linewidth,keepaspectratio]{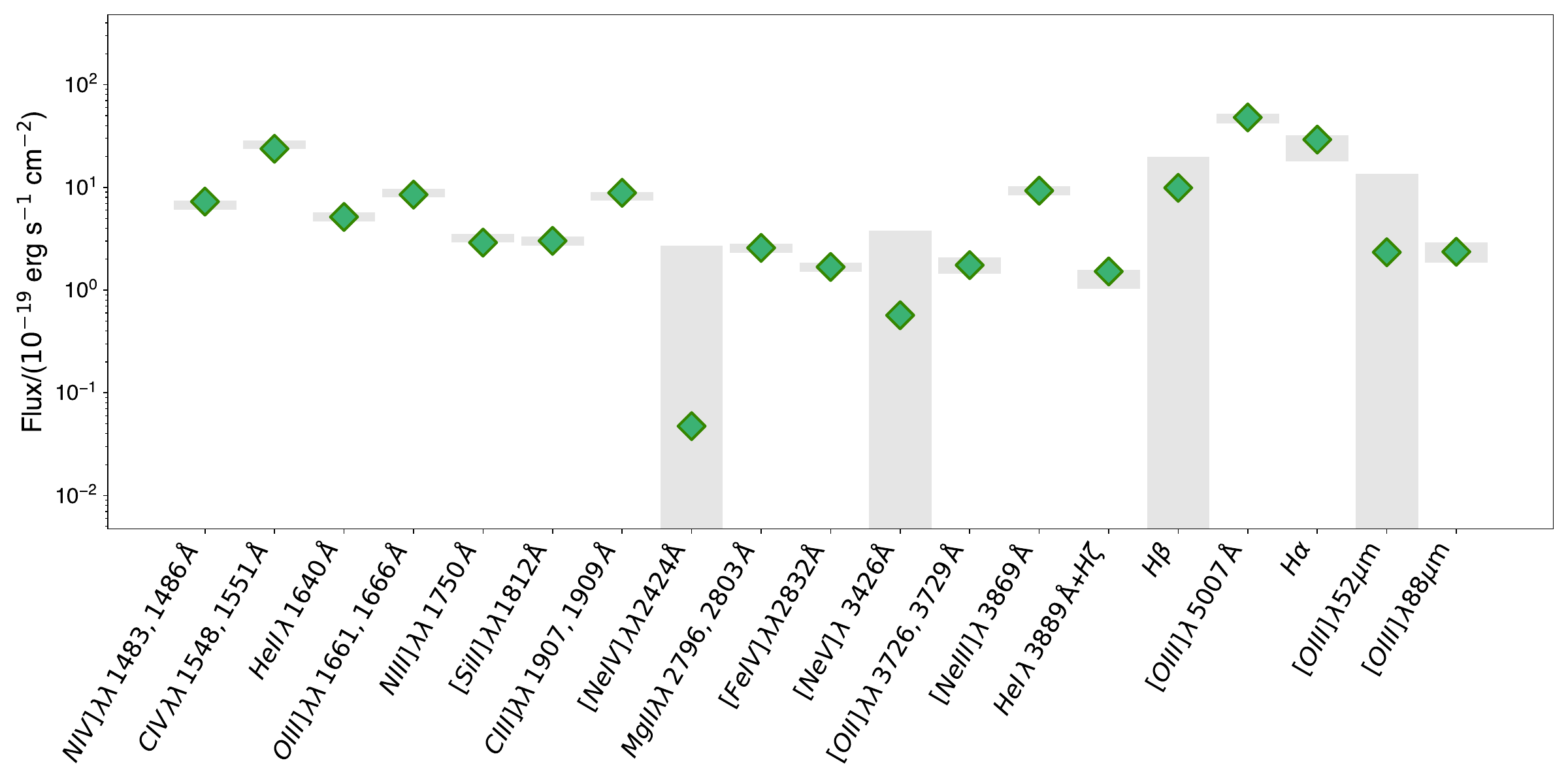}
\caption{Same as Fig.~\ref{fig_homerun_SF} but for the two components of the M2 model (radiation-bounded star-formation plus AGN). The central panel shows the relative contribution of the star-forming (red) and AGN (yellow) component to the flux of the observed lines.} 
\label{fig_homerun_AGN}
\end{figure}

\subsection{A multi-zone model of the nebular conditions with HOMERUN}\label{sec:specfit}

The emission line modeling available through \textsc{BAGPIPES} is clearly simplistic as it does not take into account the stratified nature of the ISM which is observed both at low-   \citep{Mingozzi2022} and high-redshift \citep{Ji2024}, nor the potential contribution of AGN emission. We thus performed a more refined analysis using \textsc{HOMERUN}  \citep[Highly Optimized Multi-cloud Emission-line Ratios Using photo-ionizatioN,][M24 hereafter]{Marconi2024} which allows for multi-zone modeling of emission-line spectra from photoionised regions. Briefly, \textsc{HOMERUN} assumes that the observed spectrum can be approximated by a weighted linear combination of a grid of constant-density \textsc{CLOUDY} photoionisation models \citep[][\textquote{single-cloud models}]{Ferland2013,Gunasekera2025}. 
The grid spans a range of ionization parameters and gas densities, and is computed for a fixed ionising spectrum and chemical composition. 
The weights of the single-cloud models are free parameters and are constrained by fitting the observed emission-line fluxes through a non-negative least-squares minimization of a loss function $\LF$.
For each choice of metallicity and ionizing continuum, the $U,n_{\rm e}$ grid yielding the minimum loss function value ($\LF=\LF_{\rm min}$) defines the best-fitting model, while uncertainties on the derived parameters are estimated by considering all solutions with $\LF \le \LF_{\rm min}+0.25$ (M24).
Typically, only a small fraction of the models in the $U,n_{\rm e}$ grid are assigned non-zero weights.

Single-cloud models are computed across a wide range of gas-phase metallicities, -3.3$\leq$log(Z/Z$_{\odot})\leq$+0.4, in steps of 0.2 dex, interpolated then to 0.02 dex. In addition, \textsc{HOMERUN} self-consistently accounts for non-solar abundance patterns by allowing variations in the relative abundances of other key elements (e.g., N, C, Ne, Mg), significantly improving the quality and accuracy of the fit (M24). The single-cloud models used in the following were computed using \textsc{CLOUDY} v23.01 \citep{Gunasekera2025} as described in M24 \citep[see also][]{Ceci2025,Moreschini2026}. We adopted \textsc{BPASS} stellar models \citep[][]{Eldridge2017,Stanway2018} with an upper-mass cutoff of the IMF of 300~M$_{\odot}$ as ionising continuum for the star-forming component. The narrow-line AGN ionising continuum is parametrized using a power law continuum with a fixed UV slope of $\alpha_\mathrm{UV} = -0.5$ and an exponential cut-off at a temperature $T_\mathrm{max}$ between $10^4$ and $10^7$ K (in steps of 0.5 dex), and an X-ray power-law with slope $\alpha_\mathrm{X} = -1$ with a variable X-ray-to-UV ratio $\alpha_\mathrm{ox}$ (-1.2, -1.5, -1.8) \citep{Ceci2025}. Both star-forming and AGN models are radiation-bounded by default. In addition, we consider matter-bounded star-forming models computed by stopping the photoionization calculation when the optical depth to He~II-ionising photons reaches 0.5. These models are designed to maximize the emission of He~II, which helps in modelling cases where such a strong line is observed.

\begin{figure}
\centering
\includegraphics[trim={0.cm 0.0cm 0.0cm 0.0cm},clip,width=\linewidth,keepaspectratio]{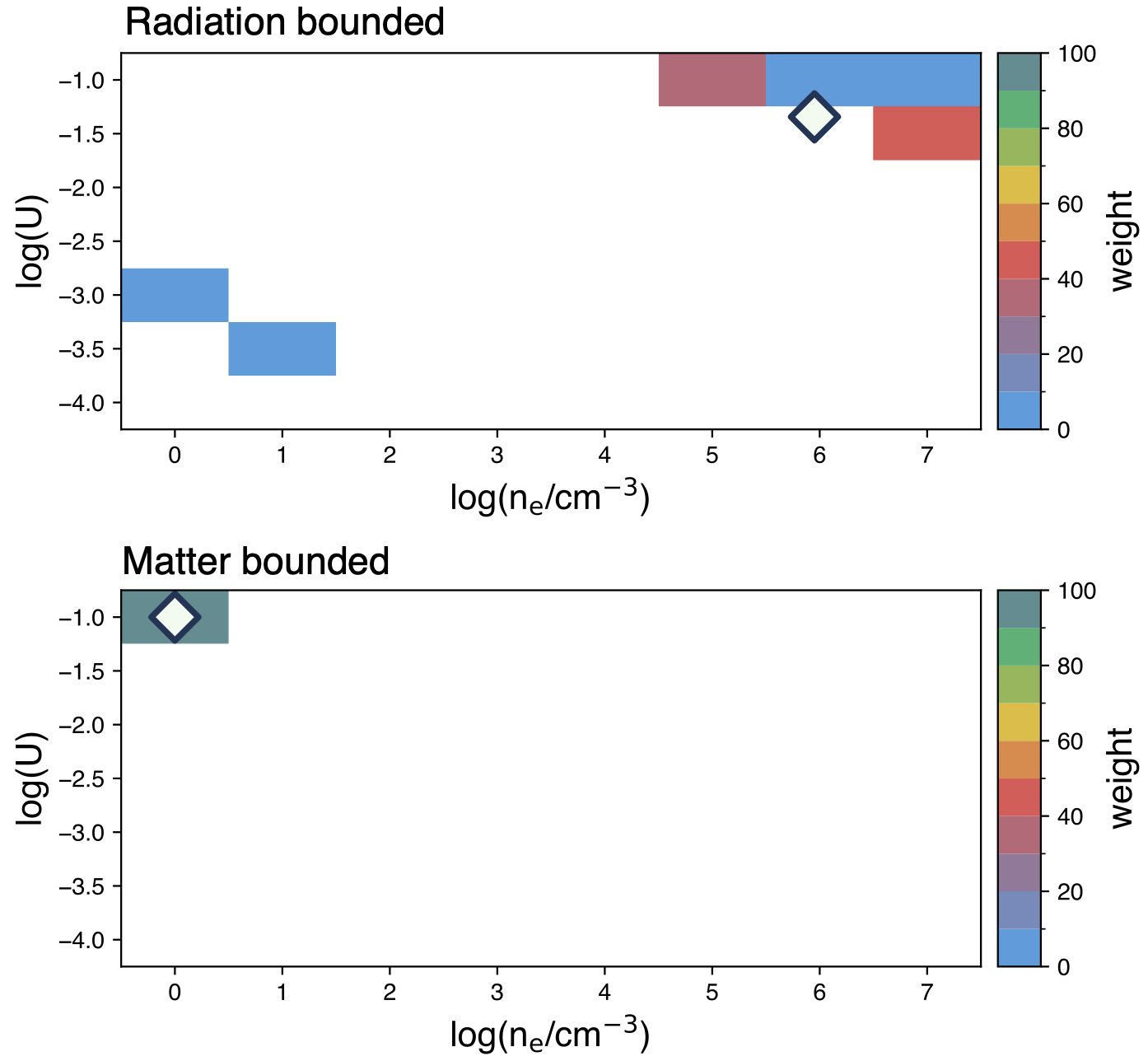}
\includegraphics[trim={0.0cm 0.0cm 0.0cm 0.0cm},clip,width=\linewidth,keepaspectratio]{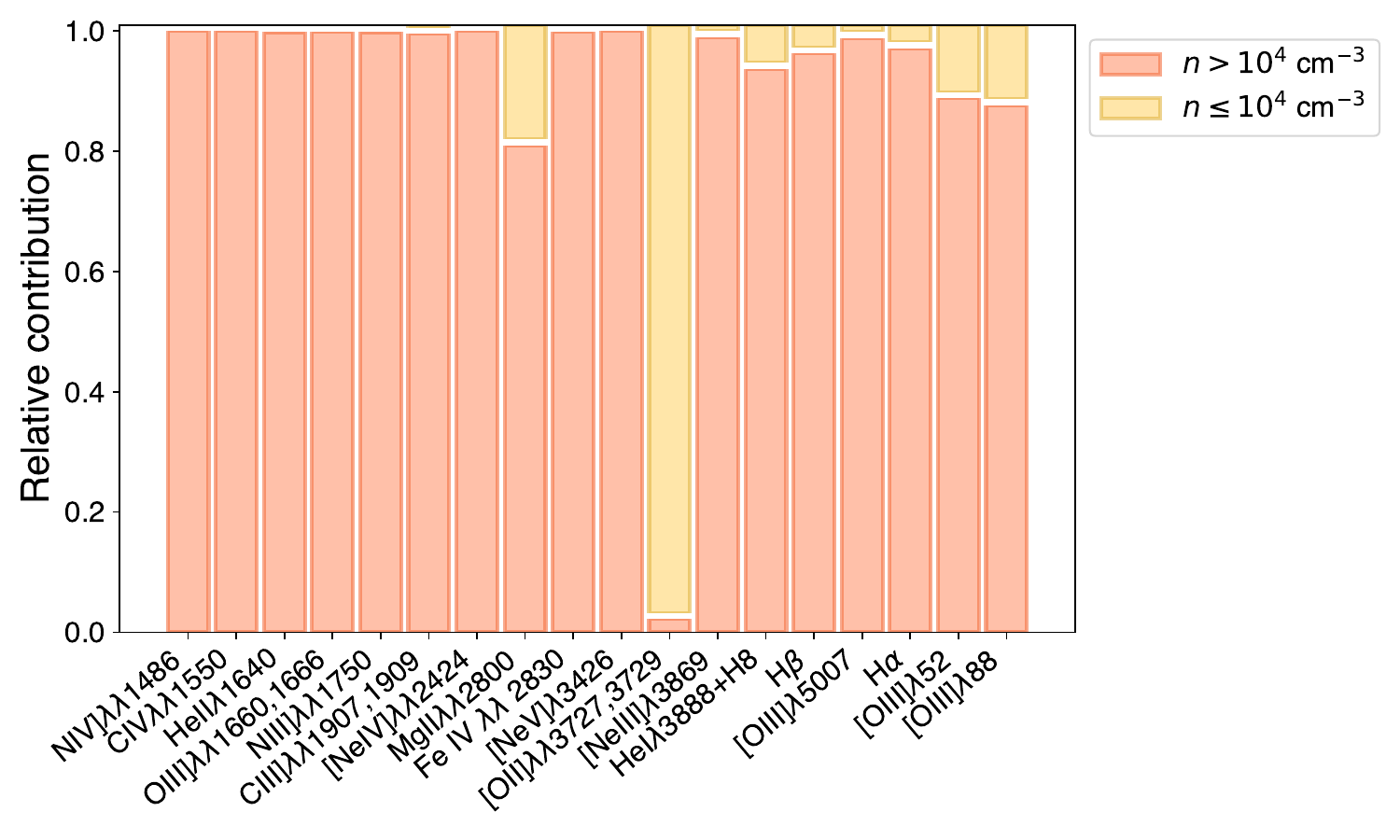}
\includegraphics[trim={0.0cm 0.0cm 0.0cm 0.0cm},clip,width=\linewidth,keepaspectratio]{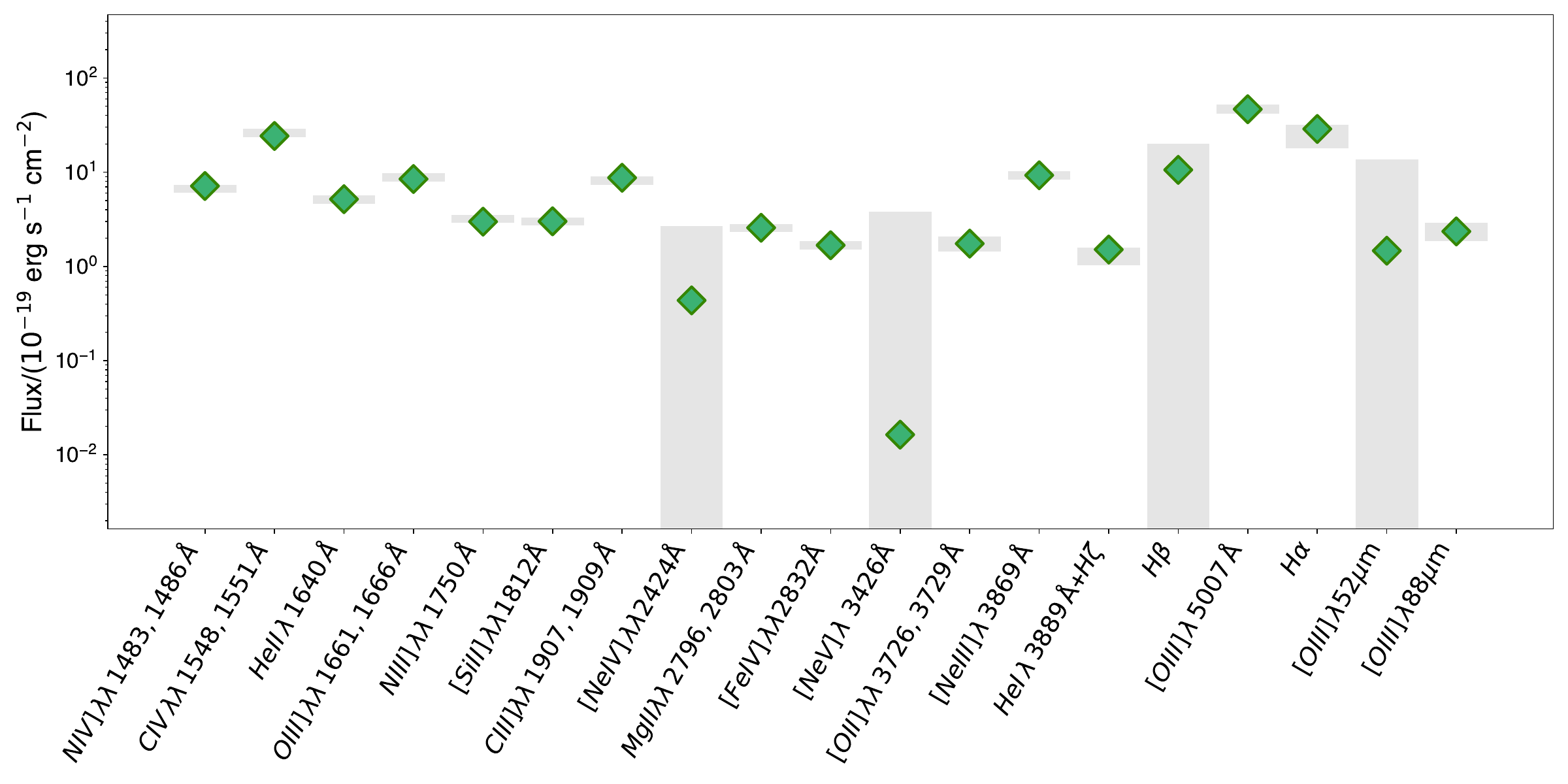}
\caption{Same as Fig.~\ref{fig_homerun_SF} but for the two components in the M3 model (radiation-bounded plus matter-bounded star-formation).} 
\label{fig_homerun_SF_radmattb}
\end{figure}

We fit NIRSpec, MIRI and ALMA lines simultaneously using three different models:
\begin{itemize}
    \item M1 - a radiation-bounded star-formation (SF) only case using grids of single-cloud models defined by $-4 \leq$log\,$U \leq - 1$, and 0$\leq$log(n$_e$/cm$^{-3})\leq$7, and spanning the ranges of metallicities and ionizing continua just described (Fig. ~\ref{fig_homerun_SF});
    \item M2 - a SF+AGN case where the SF components are radiation-bounded and grids cover the same log\,$U$ range as in the other scenarios but are limited to log(n$_e$/cm$^{-3})\leq$5, while the AGN components can have grids with $-4 \leq$log\,$U \leq 1$, and 0$<$log(n$_e$/cm$^{-3})\leq$7 (Fig. ~\ref{fig_homerun_AGN});
    \item M3 - a scenario where star formation is ongoing both in radiation-bounded and matter-bounded nebulae within the same log\,$U$ and log(n$_e$/cm$^{-3})$ ranges as in the M1 case (Fig. ~\ref{fig_homerun_SF_radmattb}).  
\end{itemize}

We find that the M1 scenario requires emission from both low- and high-density regions to match the observed lines, with a substantial contribution from the latter. The highest weight regions are found at log\,$U>-2$ and log(n$_e$/cm$^{-3})>6$, resulting in a weighted mean density log(n$_e$/cm$^{-3})$=5.41$^{+0.18}_{-0.28}$, and ionisation parameter log\,$U$=-1.45$^{+0.06}_{-0.19}$. This fit matches the observed NIRSpec, MIRI, and ALMA lines but it is unable to reproduce the high He~II luminosity which is underestimated by a factor of $\sim20$ (Fig.~\ref{fig_homerun_SF}). 

In contrast, the SF+AGN case (M2) matches all observed lines, including He~II (Fig.~\ref{fig_homerun_AGN}), predicting a weighted mean density and  ionisation parameter log(n$_e$/cm$^{-3})=3.58^{+1.42}_{-0.07}$ and log\,$U$=-1$^{+0.00}_{-0.50}$ for the SF component, and  log(n$_e$/cm$^{-3})=6.51^{+0.09}_{-1.50}$ and log\,$U$=-1.55$^{+0.55}_{-0.13}$ for the AGN one. In this scenario, the AGN powers $>$80\% of the flux from each UV line, with star-formation being the dominant contributor only for \oiiioptred~and the FIR [O~III] lines.  We note that, at variance with the M1 case, the weighted mean density of the SF component in the SF+AGN scenario is within the range estimated by \citet{Zavala2024b} from the [O~III]52$\mu$m/[O~III]88$\mu$m ratio and, similarly, log\,$U$ is in agreement with the empirical estimates provided by \citetalias{Castellano2024} and \citet{Calabro2024}.

We find that the matter-bounded SF case (M3) fits the observed constraints equally well as M2. As a consequence, this scenario cannot be distinguished from the SF+AGN one (Fig.~\ref{fig_homerun_SF_radmattb}). Similarly to the M1 case, the M3 model predicts a significant presence of high-density regions, which  contribute most of the flux of the UV lines. The high-density regions (log(n$_e$/cm$^{-3})>$4) are responsible for $>$80\% of the flux observed for each line in both the M1 and M3 scenarios, the only exception being the \oii~line, and, in the M1 case only, the FIR [O~III] lines (central panels in Fig.~\ref{fig_homerun_SF} and ~\ref{fig_homerun_SF_radmattb}).

\begin{figure}
\centering
\includegraphics[trim={0.5cm 2.5cm 1.5cm 2.5cm},clip,width=\linewidth,keepaspectratio]{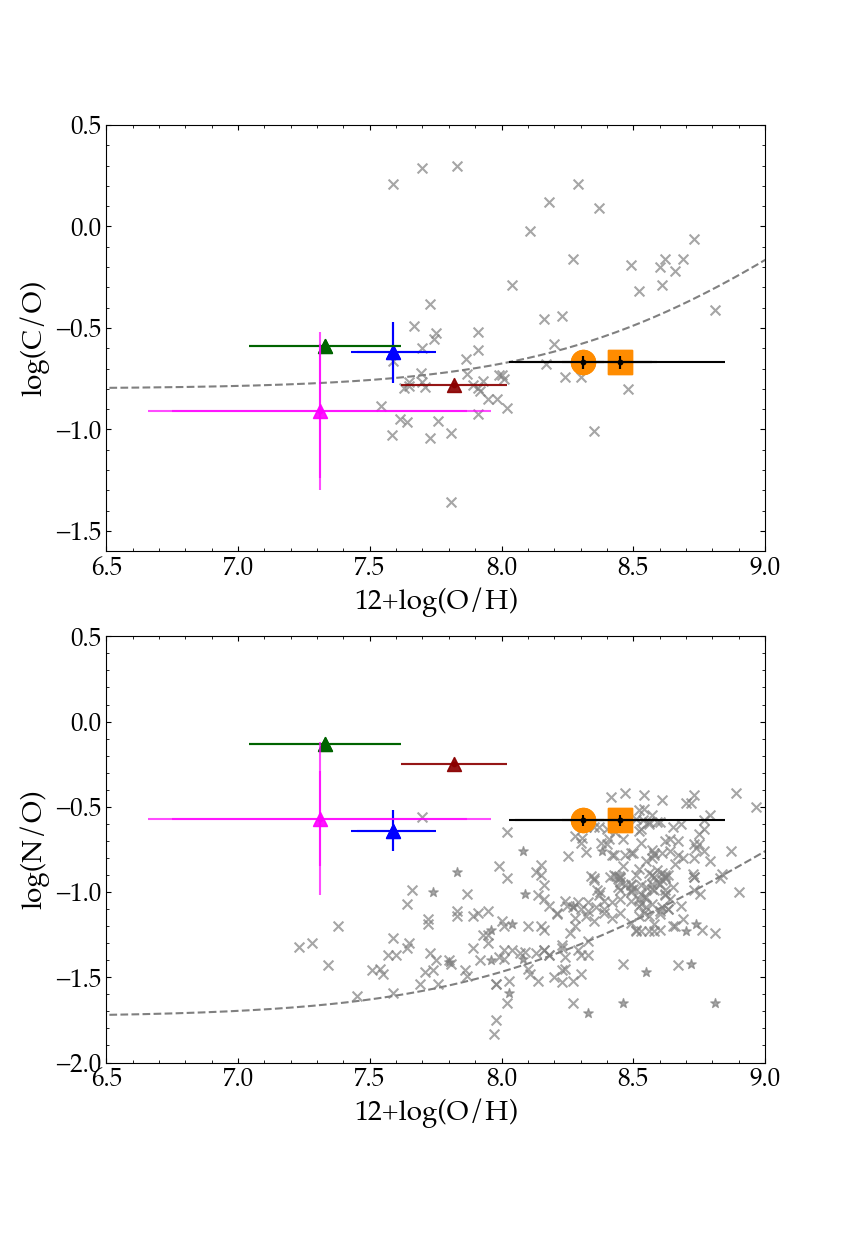}
\caption{The position of GHZ2 in the log(C/O) (top) and log(N/O) (bottom) versus 12+log(O/H) diagrams is shown as an orange filled square (circle) for the oxygen abundance estimated by \textsc{HOMERUN} in the M2 (M3) scenario. The filled triangles show reference high-redshift objects with significant Nitrogen enhancement: GNz11 \citep[][dark red]{Cameron2023}, MoM-z14 \citep[][magenta]{Naidu2025}, and the composite spectra of $z>9$ objects by \citet{Tang2025} (blue) and of $z\geq$10 CIV-strong galaxies by \citet{Roberts-Borsani2025b} (dark green). Grey crosses and stars indicate low-redshift galaxies and HII regions \citep[][]{Esteban2002,Esteban2009,Esteban2014,Senchyna2017,Berg2019,Berg2020,Izotov2023,ArellanoCordova2025} and objects at z$\sim$2-3 \citep{Rogers2026}, respectively. The scaling relations from \citet{Nicholls2017} are shown as grey dashed lines.} 
\label{fig_CNlogOH}
\end{figure}

\textsc{HOMERUN} provides a self-consistent estimate of the ionic abundances, which we find to be different from those obtained by previous works. The M2 model has 12+log(O/H)=8.45$^{+0.12}_{-0.40}$, i.e. Z/Z$_{\odot}$=0.57$^{+0.19}_{-0.34}$, significantly higher than the Z/Z$_{\odot}\sim$0.05-0.15 range indicated by \citetalias{Castellano2024} and \citet{Calabro2024} using empirical relations and strong-line diagnostics. The estimated nitrogen abundance is log(N/O)=-0.58$^{+0.02}_{-0.03}$, i.e. $\sim$2 times the solar value \citep[log(N/O)$_{\odot} = -0.86$;][]{Asplund2009}, while both carbon and iron have sub-solar abundances of log(C/O)=-0.67$^{+0.03}_{-0.03}$ and log(Fe/O) = -2.10$^{+0.12}_{-0.05}$. The star-forming models M1 and M3 yield nearly identical N, C and Fe abundances, but a slightly lower O/H ratio 12+log(O/H)=8.31$^{+0.08}_{-0.28}$ (Z/Z$_{\odot}$=0.42$^{+0.08}_{-0.20}$). 
We show in Fig.~\ref{fig_CNlogOH} the position of GHZ2 in the log(C/O) and log(N/O) versus 12+log(O/H) diagrams compared to other high-redshift sources and composite spectra showing Nitrogen enhancement, and to reference sources at lower-redshifts. Notably, the higher oxygen abundance estimated by \textsc{HOMERUN} brings GHZ2 in closer agreement with the locus populated by low- and intermediate-redshift objects with similar super-solar Nitrogen abundance, indicating that a detailed modelling of ISM conditions is fundamental to assess the origin of the N-excess in high-redshift sources.
In fact, the higher O/H found by \textsc{HOMERUN} in all considered scenarios is explained by the self-consistent treatment of emission from high-density regions which compensates the relevant higher rate of collisional de-excitation with a higher oxygen abundance. In addition, strong-line metallicity calibrations can be biased towards lower O/H values because they assume a higher [O~II] temperature than achieved in such high-density conditions (P\'{e}rez-D\'{i}az et al., in prep.). As an additional test, we determined the O/H using the direct $T_e$ method. We first measured the electronic temperature from the \oiiit/\oiiioptred~ratio using \pyneb\ \citep[\textsc{getTemDem} routine,][]{Luridiana2012, Luridiana2015}, and adopting for [OII] and [OIII] the same collisional strengths and transition probabilities used in \textsc{HOMERUN} (M24). We find 12+log(O/H)=7.29 (7.31) when assuming a constant density of log(n$_e$/cm$^{-3})=3$ ($=5$), consistent with the estimate in \citetalias{Castellano2024}, but significantly lower than the O/H derived by \textsc{HOMERUN}. The discrepancy is most likely explained by the assumption that \oiiit~and \oiiioptred~are emitted by the same region with a constant density, in agreement with the finding by \citet{Harikane2025b} that the direct $T_e$ method can significantly underestimate O/H in sources characterised by a complex density structure \citep[see also][]{Moreschini2026}.

Finally, we computed the N/O and C/O abundance ratios with \pyneb\ with the same procedure adopted in \citetalias{Castellano2024} but using the line fluxes reported in Sect.~\ref{sec:features}. We find log(N/O) $ = -0.41$ (for $n_e =10^5$~cm$^{-3}$, $T = 1.5 \times 10^4$~K) to log(N/O) $= -0.30$ (for $n_e=10^3$~cm$^{-3}$, $T=3 \times 10^4$~K), and log(C/O) $= -1.07$ (for $n_e = 10^5$~cm$^{-3}$ and $T = 1.5 \times 10^4$~K) to log(C/O) $= -0.66$ (for $n_e = 10^3$~cm$^{-3}$ and $T = 3.0 \times 10^4$~K), i.e. N/O 2.6-3.6 times the solar value and a C/O 15-40\% solar. We assessed that the lower N/O ratio compared to the \textsc{HOMERUN} estimate is due to the different atomic data used\footnote{\textsc{CLOUDY} adopts collisional strengths and transition rates from \citet{FernandezMenchero2017} (N~IV]) and \citet{Liang2012} (N~III]). \textsc{Pyneb} is based on \citet{Ramsbottom1994} and \citet{Wiese1996} for N~IV] collisional strengths and transition rates, respectively, and on \citet{Blum1992} and \citet{Galavis1998} for N~III] collisional strengths and transition rates, respectively.}.

The present analysis thus confirms the overall picture discussed in \citetalias{Castellano2024}, with GHZ2 found to be N-enhanced and carbon poor, although the measured nitrogen overabundance is lower than the one estimated in \citetalias{Castellano2024} due to the differences in line flux measurements discussed in Sect.~\ref{sec:features}, and in agreement with the abundance measured in low-redshift objects with similar O/H.

\section{Discussion}\label{sec:discussion}

The evidence emerged from the deep GHZ2 spectrum presented here, together with previous analysis based on NIRSpec, MIRI and ALMA data \citep[][]{Castellano2024,Calabro2024,Zavala2024b,Zavala2025,Mitsuhashi2025,ChavezOrtiz2025}, provides a new perspective on this puzzling source. First of all, active, compact star-formation is clearly present, as indicated by GHZ2 following the same well-established relations between [O~III]88$\mu$m luminosity and SFR, and between dynamical mass and H$\beta$ luminosity, as for star-forming galaxies and giant H~II regions \citep{Zavala2024b}.

It is also evident that star formation must be ongoing in a stratified ISM environment where both low-/intermediate-density regions and high-density ones are present, consistent with the multizone density structure inferred for similar high-redshift sources \citep[][]{Harikane2025b,Topping2025b}. The presence of star formation at 100 $ < $ n$_e$/cm$^{-3} < $ 4000 is indicated by the constraints on electron density from ALMA data \citep{Zavala2024b}, while our analysis in Sect.~\ref{sec:specfit} shows that this is not enough to reproduce the UV spectrum, but regions with log(n$_e$/cm$^{-3})\gtrsim$4 are needed, even when including an AGN contribution, and regardless of the radiation- or matter-bounded nature of the star-forming regions. 
The detection of variability in the \oiiibowen~fluorescence line (Sect.~\ref{subsec:Bowen}) strongly points to a more complex scenario, where AGN emission adds to compact star formation in a stratified ISM to produce the peculiar emission landscape of GHZ2. In fact, the short time span between the two subsets of our spectroscopic observations implies a coherent change in a region of less than 0.02 pc, consistent with the size of a broad-line region (BLR). The gas in the BLR can reach the extreme densities and the high level of ionising flux needed to generate this fluorescence line, which is indeed observed in local \citep{Schachter1990} and distant \citep{Lanzuisi2015} AGN, including short-term variable ones \citep{Makrygianni2023}. 
This finding, together with the good fit of the observed spectra obtained for the SF+AGN case with \textsc{HOMERUN}, as well as the similar results obtained with \textsc{BEAGLE-AGN} by \citet{ChavezOrtiz2025} and with nitrogen-enhanced photoionisation models by \citet{Zhu2025}, provide strong support to a composite nature of GHZ2. The observed He II flux is most naturally accounted for by an AGN contribution, unless one invokes the presence of additional, highly ionizing sources not considered in current star-formation models, such as very massive stars \citep[e.g.][]{Vink2023}, super-massive stars \citep[e.g.][]{Denissenkov2014}, or Population III stars \citep[e.g.][]{Nakajima2022,Wang2024}. The presence of matter-bounded star-forming regions can also explain the He~II luminosity, but it is apparently at odds with the high column density of neutral gas measured from the Ly$\alpha$ damping wing profile, assuming the DLA originates from dense ISM gas on $\sim$100 pc scales \citep[][]{Gelli2025}. 

The combined properties of GHZ2 point to this source being consistent with the scenario proposed by \citet{Isobe2025} drawing a connection between compact star formation, N-enhancement, and AGN activity. In such a scenario, dense gas in the nuclear region of a galaxy forms stars efficiently, leading to the ISM being polluted by the N-enriched gas ejected by massive stars \citep[see also][]{Kobayashi2024,Topping2025,Naidu2025,Morel2025}. The massive stars in conditions of sufficient gas accretion would first develop intermediate-mass black holes which eventually evolve into SMBHs. This scenario may explain the high N/O found in z=4-7 AGN by \citet{Isobe2025}, as well as the large number of N-enriched sources at high redshift with AGN signatures, such as GN-z11 \citep{Maiolino2023} and GHZ9 \citep{Napolitano2024b}. Finally, the N-rich gas trapped in the high-pressure clouds of the nuclear regions \citep[][]{Pascale2023} may eventually lead to the formation of the N-enriched stellar populations of present-day globular clusters \citep[e.g.,][]{Charbonnel2023,Senchyna2023,DAntona2023,Marques-Chaves2024,Ji2025}. The remarkable compactness \citep[$\lesssim$100 pc,][]{Yang2022b,Ono2023}, and high $\Sigma_{\rm SFR}$ (log($\Sigma_{\rm SFR}$~[M$_{\odot}$~yr$^{-1}$~kpc$^{-2}$]) $\approx 1.9$) of GHZ2 are in line with this scenario, similarly to other compact, highly ionising N-emitters at similar redshifts \citep{Harikane2025,Naidu2025}. In fact, a high $\Sigma_{\rm SFR}$ possibly due to a burst of star formation is a common property of compact, high-redshift objects with high N/O \citep[][]{Topping2025,Tang2025}. This scenario is also supported by the large N(H~I) in front of GHZ2 which is suggestive of a rapid gas accretion that feeds compact star-formation, and possibly AGN growth.

\section{Summary and future prospects}\label{sec:summary}

We presented the analysis of GHZ2 at z=12.3 exploiting 9.1 hours of NIRSpec PRISM observations, more than doubling the exposure time analysed in \citetalias{Castellano2024}. The final GHZ2 spectrum is publicly released together with all reduced spectra and redshift measurements of the Cycle 2 program GO-3073 as described in Sec.~\ref{sec:appendix-datarelease}. The analysis of the full-depth NIRSpec dataset enabled us to measure at high significance the UV emission lines detected in the first-epoch observations and to find additional emission and absorption features   (Table.~\ref{tab:lines}). The O~III 3133~\AA\ line emitted via Bowen resonance fluorescence is detected in the first epoch observations only, indicating variability on a rest-frame time scale of only $\sim$19 days (Sec.~\ref{subsec:Bowen}), consistent with its emission originating from dense gas of a broad-line region as in other known AGN at lower redshfits. Finally, the analysis of the damping wing profile of GHZ2 indicates that GHZ2 is embedded in a neutral gas component with a column density of log~N(HI)/[cm$^{-2}$]= 22.35$\pm$0.37 (Sec.~\ref{subsec:dampingwing}). 

We analysed the NIRSpec observations in conjunction with the available MIRI and ALMA constraints with the \textsc{HOMERUN} code (Sec.~\ref{sec:specfit}).
The multi-zone modeling shows that GHZ2 is consistent with both star-formation only and star-formation plus AGN models . However, the bright He~II line (EW$\sim$8\AA) requires either the presence of both radiation- and matter-bounded star-forming clouds, or AGN emission. Under all considered scenarios, star formation must be ongoing in a stratified ISM where both low-/intermediate- density regions and high-density ones (log(n$_e$/cm$^{-3})\gtrsim$4) are present.  \textsc{HOMERUN} also provides self-consistent estimates of element abundances. Regardless of the sources of ionising photons, GHZ2 is found to be N-enhanced (log(N/O)=-0.58$^{+0.02}_{-0.03}$, i.e., $\sim$2 times the solar value), and with sub-solar C and Fe abundances. Instead, once the high-density ISM is taken into account, the oxygen abundance is estimated to be significantly higher than the one based on empirical relations, ranging from $\sim$0.4 \zzsun~in the star-forming case to $\sim$0.6 \zzsun~for the AGN+SF scenario.

The detection of variability in the \oiiibowen~fluorescence line and the \textsc{HOMERUN} analysis provide support to a composite star-forming plus AGN nature of GHZ2, consistently with recent results \citep{ChavezOrtiz2025,Zhu2025}. The properties of GHZ2 are in line with the scenario proposed by \citet{Isobe2025} in which dense gas feeds both AGN accretion and nuclear star-formation which eventually pollutes the ISM with N-enriched gas (Sec.~\ref{sec:discussion}). 

Unfortunately, several ingredients remain poorly constrained. In particular, it is not yet possible to directly measure the relative contributions of star formation and AGN activity, nor to determine the density structure of the ISM. 

The available deep NIRSpec PRISM and MIRI observations lack the resolving power to investigate in more detail the nature of GHZ2. Further progress will be enabled by the forthcoming JWST plus ALMA Cycle 4 program GO-7201, which will constrain the ISM structure by resolving density-sensitive UV line doublets with the R$\sim$2700 gratings and through deep ALMA Band 8 observations to tighten the constraints on the [O~III]52$\mu$m line. This will also be fundamental to confirm the relatively high oxygen abundance estimated by our multi-zone modeling. The upcoming MIRI MRS R$\sim$3000 observations of GHZ2 under program GO-7078 can detect broad components of the H$\alpha$ line, hence directly measuring the BLR fraction. The MIRI imaging obtained under program GO-9165 will yield tighter constraints on the SFR burstiness and stellar mass. Meanwhile, ALMA is targeting the [O~I]145$\mu$m, [C~II]158$\mu$m (program 2024.1.00536.S) and [N~III]57$\mu$m (program 2024.1.01645.S) lines of GHZ2 to probe its cold ISM and constrain chemical abundances. The combined power of JWST and ALMA has the potential to explore the complexity of GHZ2 and similar high-redshift objects. However, in the long term, it will also be crucial to fill the gap between the UV and optical rest-frame accessible by JWST, and the FIR observed by ALMA. By targeting the near- and mid-infrared rest-frame, PRIMA \citep[][]{Glenn2025} will be able to observe high ionisation emission lines that are powerful tracers of AGN contribution and chemical composition   \citep[e.g.][]{Tommasin2010,PerezDiaz2022}.

Only a multi-instrument follow-up approach will unveil the constituents of GHZ2 and other similar objects at cosmic dawn, which can serve as \textquote{Rosetta stones} for the first phases of galaxy and AGN assembly.

\section*{Acknowledgments}
We  thank  the  referee  for  the  constructive comments that helped us improve the manuscript. We thank S. Finkelstein, Y. Harikane, C. Mason, and D. Stark for the useful comments.
We thank Tony Roman (Program Coordinator) and Glenn Wahlgren (NIRSpec reviewer) for the assistance in the preparation of GO-3073 observations.  
This work is based on observations made with the NASA/ESA/CSA {\it James Webb Space Telescope (JWST)}. The JWST data presented in this article were obtained from the Mikulski Archive for Space Telescopes (MAST) at the Space Telescope Science Institute. The specific observations analysed are associated with program JWST-GO-3073 and can be accessed via \url{https://doi.org/10.17909/4r6b-bx96} (first pointing) and \url{https://doi:10.17909/zq4g-r525} (second pointing).
We acknowledge financial support from NASA through grant JWST-ERS-1324 and JWST-GO-3073. Support was also provided by the PRIN 2022 MUR project 2022CB3PJ3 – First Light And Galaxy aSsembly (FLAGS) funded by the European Union – Next Generation EU, by INAF GO Grant 2024 "Revealing the nature of bright galaxies at cosmic dawn with deep JWST spectroscopy", by INAF Mini-grant 2022 ``Reionization and Fundamental Cosmology with High-Redshift Galaxies", and by INAF Large Grant 2022 “Extragalactic Surveys with JWST”.  L.N. acknowledges support from grant ``Progetti per Avvio alla Ricerca - Tipo 1, Unveiling Cosmic Dawn: Galaxy Evolution with CAPERS" (AR1241906F947685). EV acknowledges financial support through grants INAF GO Grant 2024 ``Mapping Star Cluster Feedback in a Galaxy 450 Myr after the Big Bang'' and by the European Union – NextGenerationEU within PRIN 2022 project n.20229YBSAN - Globular clusters in cosmological simulations and lensed fields: from their birth to the present epoch.
AM acknowledges support from project PRIN-MUR project “PROMETEUS” financed by the European Union - Next Generation EU, Mission 4 Component 1 CUP B53D2300475000.
AM acknowledges support from Ricerca Fondamentale INAF under Mini Grant 2023 "Quantitative Spectroscopy of Ionized Nebulae and Galaxies (QSING)" and  
under Data Analysis Grant 2024 “Accurate measurements of metallicity in galaxies with a new approach to photoionization modelling”.\\

\bibliographystyle{mn2e} 

\begin{appendix}
\section{Full data release of the GO-3073 NIRSpec PRISM spectra}\label{sec:appendix-datarelease}

The GO-3073 NIRSpec observations of the GLASS-JWST field comprise two partially overlapping pointings, observed on 2023 Oct. 24 with aperture position angle (APA) of 175 deg, and on 2024 July 3-4 with APA=30 deg., respectively. Each pointing is divided into three visits of exposure time of 6567~s each, adopting a NRSIRS2 readout pattern, standard 3-shutter ``slits'', and a 3-point nodding. 
The main targets of the program were GHZ2 and the five colour-selected z$\sim$10 candidates within the GLASS-JWST ERS region discussed in \citet{Castellano2023a}. Additional colour-selected candidates \citep[][]{Castellano2022b,Atek2023,Harikane2022b,McLeod2024}, and objects with photometric redshift z $\gtrsim$ 9 from \citet{Merlin2024} were included as high-priority fillers, for a total of 27 candidates at z$\sim$9-12. The allocation of the remaining slitlets was maximised with the MSA planning tool to target sources from the \citet{Paris2023} and \citet{Merlin2024} catalogues, assigning a higher priority to objects with photometric redshift z$_{phot}\geq$5. A total of 957 primary 
targets have been observed by the GO-3073 program.  

The data were reduced as outlined by \cite{ArrabalHaro2023a} and \cite{ArrabalHaro2023b} with the STScI Calibration Pipeline\footnote{\url{https://jwst-pipeline.readthedocs.io/en/latest/index.html}} version 1.13.4, and the Calibration Reference Data System (CRDS) mapping 1197. The pipeline modules are divided into three components. In summary, the \textsc{calwebb\_detector1} module corrects for detector $1/f$ noise, subtracts dark current and bias, and generates count-rate maps (CRMs) from the uncalibrated images. The \textsc{calwebb\_spec2} module creates two-dimensional (2D) cutouts of the slitlets, corrects for flat-fielding, performs background subtraction using the three-nod pattern, executes photometric and wavelength calibrations, and resamples the 2D spectra to correct distortions of the spectral trace. The \textsc{calwebb\_spec3} module combines images from the three nods, utilising customized extraction apertures to extract the one-dimensional (1D) spectra. 

\begin{figure}[ht]
\centering
\includegraphics[trim={0.cm 0.cm 0.cm 0.cm},clip,width=\linewidth,keepaspectratio]{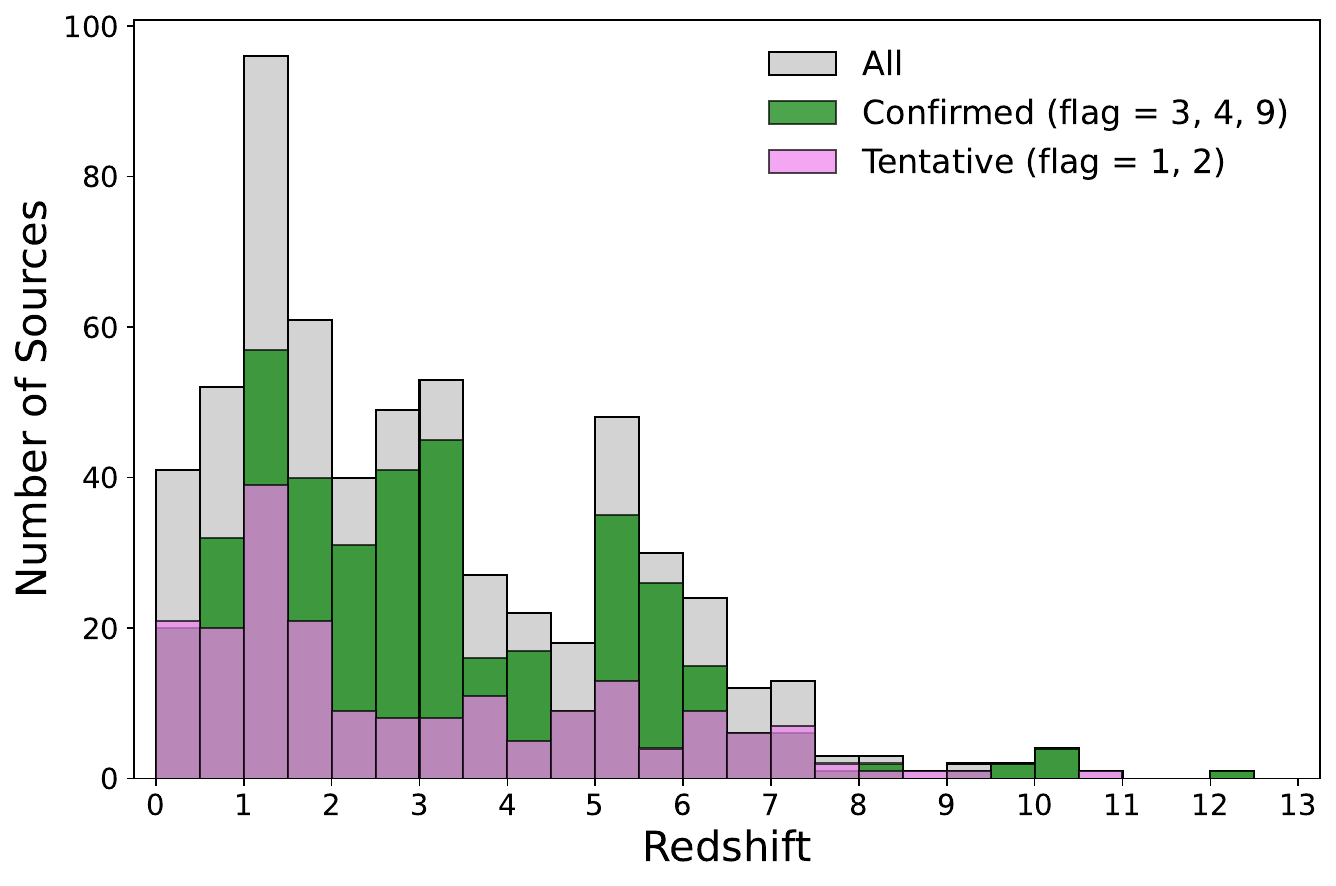}
\includegraphics[trim={0.cm 0.cm 0.cm 0.cm},clip,width=\linewidth,keepaspectratio]{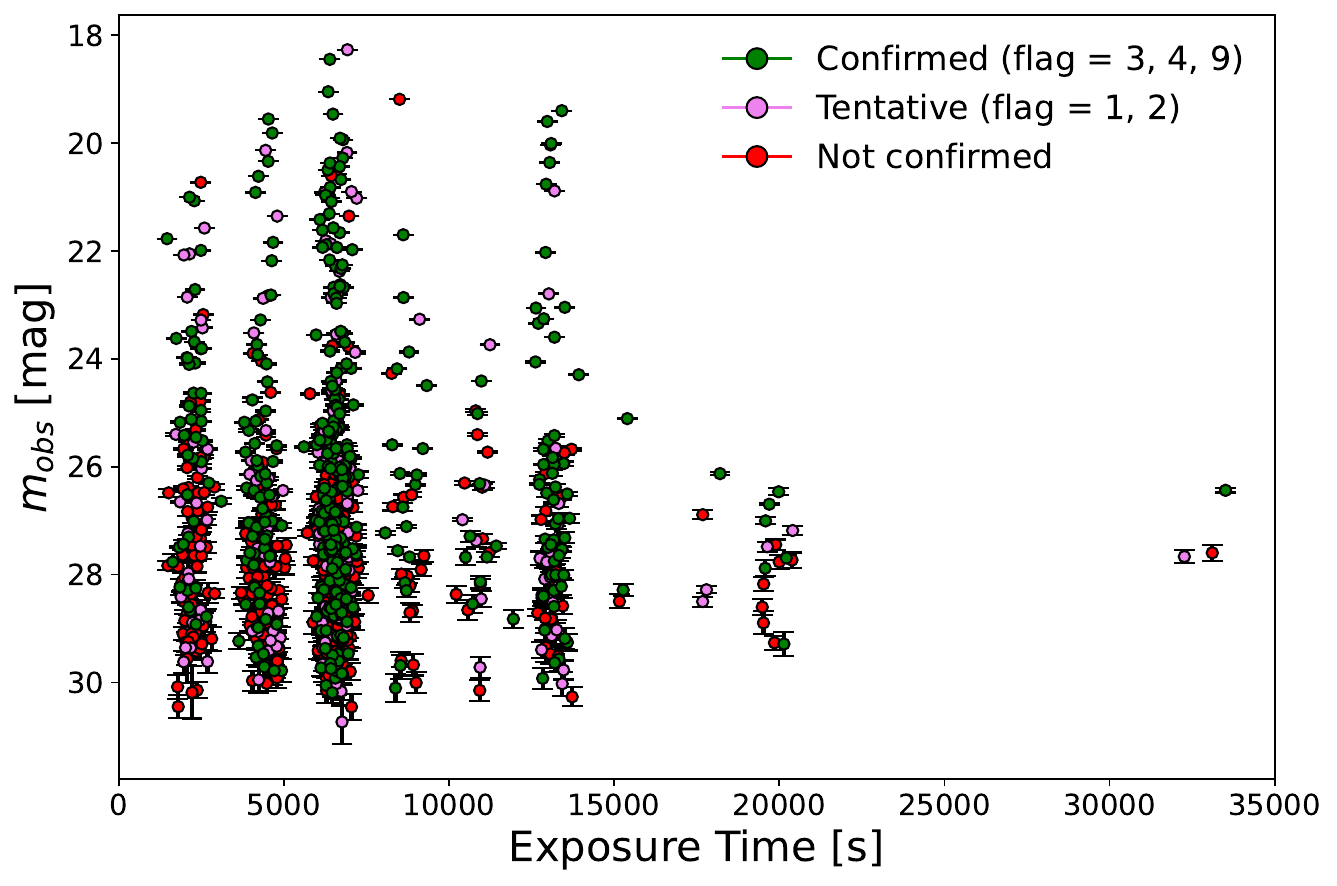}
\caption{\textbf{Top:} redshift distribution of all the targets spectroscopically confirmed by program GO-3073. \textbf{Bottom:} observed magnitude in the F444W band versus total exposure time of the confirmed objects colour-coded according to the relevant spectroscopic quality flag. Data points have been slightly shifted in the exposure time axis for clarity.} 
\label{fig_datarelease}
\end{figure}

\begin{table*}
    \centering
        \caption{Example of the spectroscopic redshift information from the GO-3073 program.}\footnote{ID$_{\mathrm{ASTRODEEP}}$ and ID$_{\mathrm{Paris23}}$ are the identification numbers from \citet{Merlin2024} and \citet{Paris2023}, respectively. The object coordinates are from \citet{Paris2023} when MPT\_ID=ID$_{\mathrm{Paris23}}$, and from \citet{Merlin2024} otherwise.}
    \label{tab:summary_data}
    \begin{tabular}{ccccccccc}
    \hline \hline
    \noalign{\smallskip}
    \text{MPT\_ID} & \text{RA} \ [\text{deg}] & \text{DEC} \ [\text{deg}] & z$_{\mathrm{spec}}$ & \text{Flag} & t$^{\mathrm{exp}}$ \ [\text{s}] & ID$_{\mathrm{ASTRODEEP}}$ & ID$_{\mathrm{Paris23}}$ & Notes\\
    \noalign{\smallskip}
    \hline
    \noalign{\smallskip}
    \noalign{\smallskip}
    22600 & 3.4989827 & -30.3247534 & 12.34 & 4 & 32825 & 39462 & 22600 & GHZ2\\
  22619 & 3.492319 & -30.3214661 & 2.86 & 3 & 6565 & 37632 & 22619 &\\
  22626 & 3.5036937 & -30.3179746 & 6.28 & 4 & 19695 & 36846 & 22626 &\\
  22643 & 3.4567012 & -30.3225594 & - & 0 & 2188 & 37915 & 22643 &\\
  22650 & 3.4565751 & -30.322097 & 1.75 & 1 & 6565 & 39798 & 22650 &\\
  22706 & 3.4983802 & -30.3222831 & 8.20 & 3 & 13130 & 37113 & 22706 &\\
    \noalign{\smallskip}
    \noalign{\smallskip}
    \hline
    \hline
    \end{tabular}
\end{table*}

At this stage we apply an automatic tool to the pipeline 2D spectra to search for possible interlopers that are spatially offset from the main target, as they could affect the standard three-nod background subtraction of the primary sources. Briefly, our automatic tool searches for obvious continuum traces in the spatial direction of the 2D spectra. After summing the 2D spectral information in the wavelength axis for each spatial pixel, we normalize the array of integrated fluxes using its maximum. We then filter out negative values associated to spatial pixels impacted by negative traces caused by background subtraction, obtaining an array of normalized non-negative signal and corresponding error as a function of spatial pixels. To consider an aggregated pixel region as a robust source trace candidate, we only consider group of pixels whose signal exceeds a threshold set by the median value of the array. We then flag local maxima on each identified region. These values are used as a starting point for a spatial Gaussian fit on the pixel axis, which describes the source trace. Traces associated to a spatial extension which is less then the expected instrument PSF are discarded. The outputs from the automatic tool and fitting routine are visually inspected. Overall, we identify 132 2D spectra with a serendipitous source in the slit which were already known from catalogs, 22 2D spectra with an unexpected serendipitous companion in the slit, and 88 2D spectra which present a spatially offset spurious artifact that would have negatively impacted the quality of the primary target extraction. For all these cases we apply a custom background subtraction to isolate and take away any dither that would negatively affect the standard three-nod background subtraction of the primary sources. In the cases with a serendipitous secondary source, we extract both the primary target and the interloper. 

In total, we extract 979 targets. As described in \cite{Napolitano2025}, the customized extraction apertures defined by the spatial Gaussian fit resulted in a increase of the average SNR ratio on the 1D spectra by up to a factor of 1.25 compared to the standard MSA extraction apertures.

The redshift measurements have been performed independently by three team members visually inspecting 1D and 2D spectra and manually matching the spectral position of emission and absorption features. The reliability of the redshift measurements was evaluated using the same flagging convention as for previous surveys such as VVDS \citep{LeFevre2015}, zCOSMOS \citep{Lilly2007} and VANDELS \citep{Pentericci2018,Garilli2021}. Namely, Flag=3 and =4 objects are highly reliable redshift measurements determined from two or more clear spectral features, with the highest flag indicating spectra with high SNR across the entire wavelength range; Flag=9 is assigned to objects with a reliable redshift determined from a single clear spectral feature; Flag=2 measurements are supported by continuum shape and one or more low SNR features, and based on previous experience have a $\sim$75\% chance of being correct. A Flag=1 is assigned to redshift identifications which considers only  a single tentative spectroscopic feature.

We show in Fig.~\ref{fig_datarelease} the redshift distribution of the confirmed and tentative targets (top panel), and their position in the observed magnitude versus total exposure time plane (bottom panel). 
We find 407 objects with Flag$\geq$3, i.e. highly reliable confirmed measurements, and 196 tentative redshifts with 1$\leq$Flag$\leq$2. A redshift could not be determined for a total of 376 objects, which are assigned Flag=0 in the catalog. The latter category mostly comprises objects with short exposure times and faint targets (bottom panel in Fig.~\ref{fig_datarelease}). In fact, we find a high success rate for the spectroscopic confirmation of bright sources ($m_{obs}\lesssim$26), ranging from $\sim$60\% with an exposure time $t_{exp}\lesssim$1 hr to $\sim$80\% with $t_{exp}\gtrsim$3 hrs. The spectroscopic confirmation rate is significantly lower for faint  objects with $m_{obs}>$26, from $\lesssim$20\% ($t_{exp}\lesssim$1 hr) to $\sim$40\% ($t_{exp}\gtrsim$3 hrs).

The reduced 1D and 2D spectra and the redshift measurements are publicly available at the ASTRODEEP website\footnote{\url{http://www.astrodeep.eu/go3073/}}. We show an example of the available information in Table~\ref{tab:summary_data}.

The data release includes the final spectra of GHZ2, of the z=10.145 X-ray AGN GHZ9 \citep{Napolitano2024b}, and of the other six confirmed objects at z$>$9.5 presented in \citet{Napolitano2025}. Other confirmed targets include the galaxy members of the A2744 cluster presented by \citet{Vulcani2025}, and objects at z$\geq$4 discussed in \citet{Roberts-Borsani2024}, \citet{Roberts-Borsani2025b}, \citet{Llerena2025}, and \citet{Morishita2025}.

\section{Additional information on the GHZ2 NIRspec spctrum}\label{sec:appendix-addinfo}

\begin{figure}[!ht]
\centering
\includegraphics[trim={0.cm 0.cm 0.cm 0.cm},clip,width=0.9\linewidth,keepaspectratio]{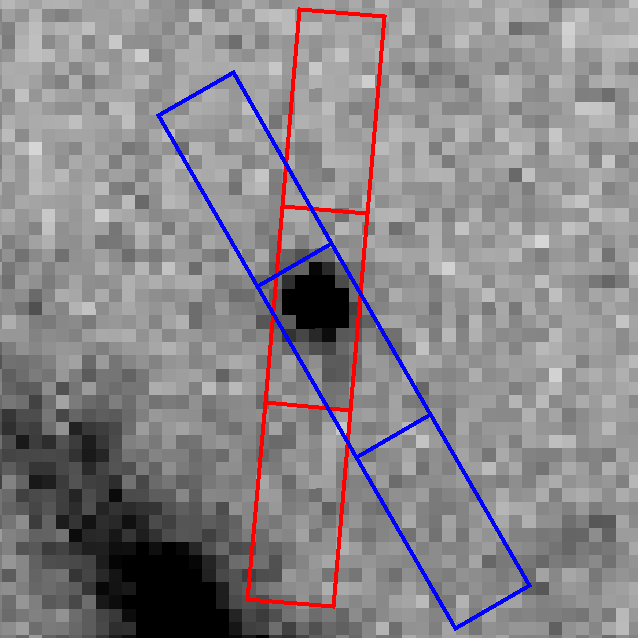}
\caption{The three-shutter slitlets of the first (red, APA=175 deg.) and second (blue, APA=30 deg.) epoch observations of GHZ2 overlayed on a 1.5$\times$1.5 arcsec snapshot of the F200W NIRCam image of the GLASS-JWST field.} 
\label{fig_GHZ2appendix}
\end{figure}

We present additional information on the observations of GHZ2. The position of the NIRSpec slitlets in the two pointings of program GO-3073 are shown in Fig.~\ref{fig_GHZ2appendix}.

\begin{figure*}[ht]
\centering
\includegraphics[trim={0.3cm 3.5cm 0.55cm 0.5cm},clip,width=0.23\linewidth,keepaspectratio]{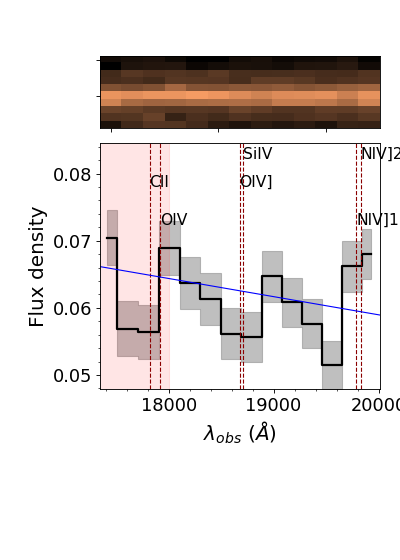}
\includegraphics[trim={0.3cm 3.5cm 0.55cm 0.8cm},clip,width=0.23\linewidth,keepaspectratio]{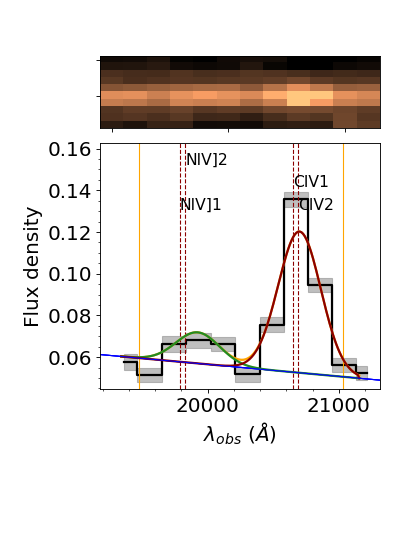}
\includegraphics[trim={0.3cm 3.5cm 0.55cm 0.8cm},clip,width=0.23\linewidth,keepaspectratio]{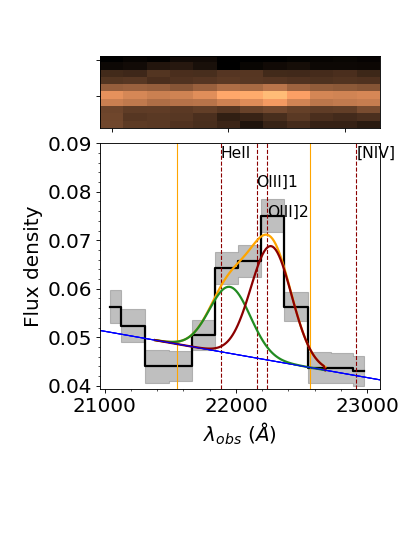}
\includegraphics[trim={0.3cm 3.5cm 0.55cm 0.8cm},clip,width=0.23\linewidth,keepaspectratio]{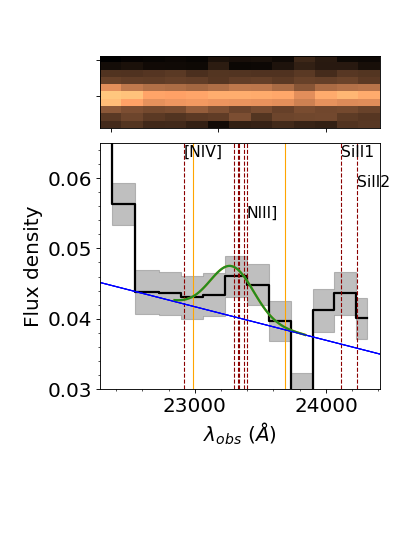}
\includegraphics[trim={0.3cm 3.5cm 0.55cm 0.8cm},clip,width=0.23\linewidth,keepaspectratio]{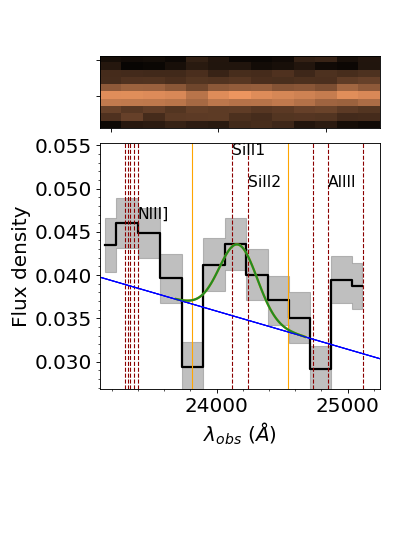}
\includegraphics[trim={0.3cm 3.5cm 0.55cm 0.8cm},clip,width=0.23\linewidth,keepaspectratio]{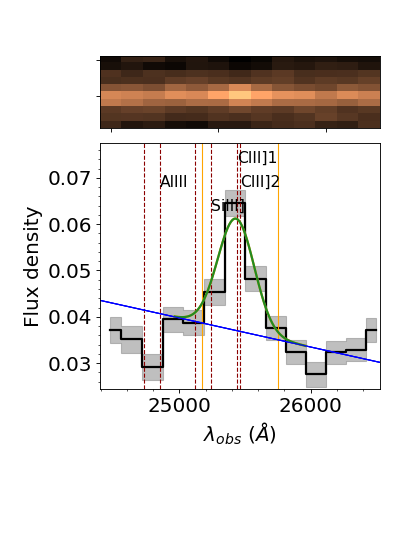}
\includegraphics[trim={0.3cm 3.5cm 0.55cm 0.8cm},clip,width=0.23\linewidth,keepaspectratio]{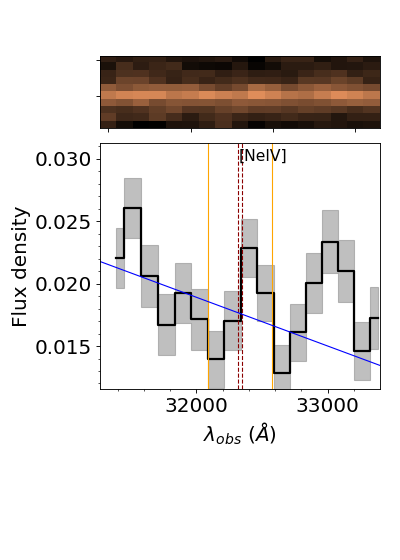}
\includegraphics[trim={0.3cm 3.5cm 0.55cm 0.8cm},clip,width=0.23\linewidth,keepaspectratio]{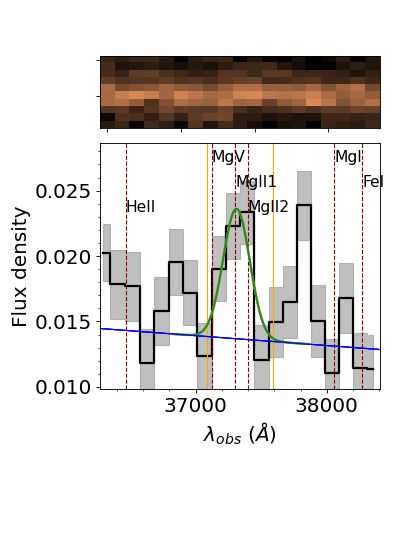}
\includegraphics[trim={0.3cm 3.5cm 0.55cm 0.8cm},clip,width=0.23\linewidth,keepaspectratio]{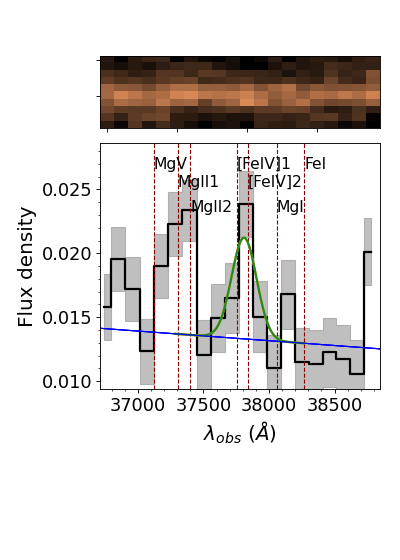}
\includegraphics[trim={0.3cm 3.5cm 0.55cm 0.8cm},clip,width=0.23\linewidth,keepaspectratio]{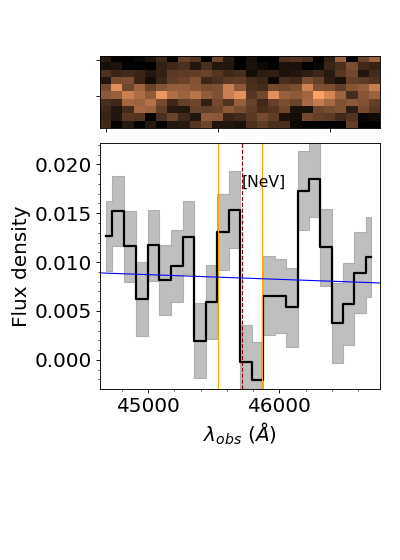}
\includegraphics[trim={0.3cm 3.5cm 0.55cm 0.8cm},clip,width=0.23\linewidth,keepaspectratio]{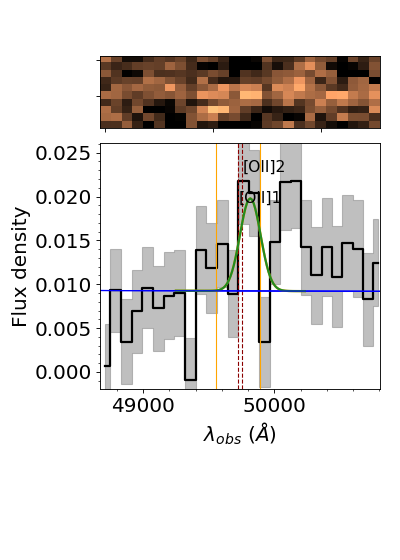}
\includegraphics[trim={0.3cm 3.5cm 0.55cm 0.8cm},clip,width=0.23\linewidth,keepaspectratio]{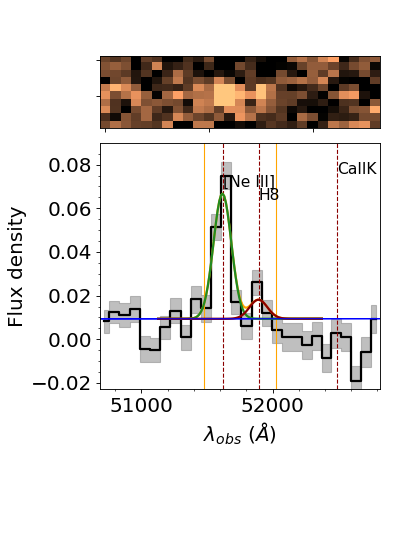}
\caption{From top left to bottom right: snapshots of the NIRSpec spectrum in regions with width of 160~\AA\ rest-frame centered at the position of: \sIivoiv~absorption, \civ\ and \niv, \heii\ and \oiii, \niii, \sIii, \ciii, \neiv, \mgii, \feiv, \nev, \oii, \neiii~and \hzetahei. All flux densities are in units of 10$^{-19}$ \ergsA. The gray shaded area shows the 1$\sigma$ uncertainty in each pixel. Red dashed lines indicate the wavelength of all potential features in the relevant spectral range. The vertical orange lines enclose the region where the signal-to-noise ratio (SNR) of the feature is evaluated from direct integration. For all significant emission lines the relevant single-Gaussian fit is shown in green. When a double-Gaussian fit was used, the two components are shown as green and red curves, and the sum of the two in orange. The blue line in each panel shows the estimated UV continuum. In the first panel we highlight as a red shaded region the spectral range that has been masked in the present analysis due to background subtraction potentially contaminated by the H$\alpha$ line of a secondary target.} 
\label{fig_lines}
\end{figure*}

\begin{figure*}[ht]
\centering
\includegraphics[trim={0.cm 0.cm 0.cm 0.cm},clip,height=0.2\linewidth,keepaspectratio]{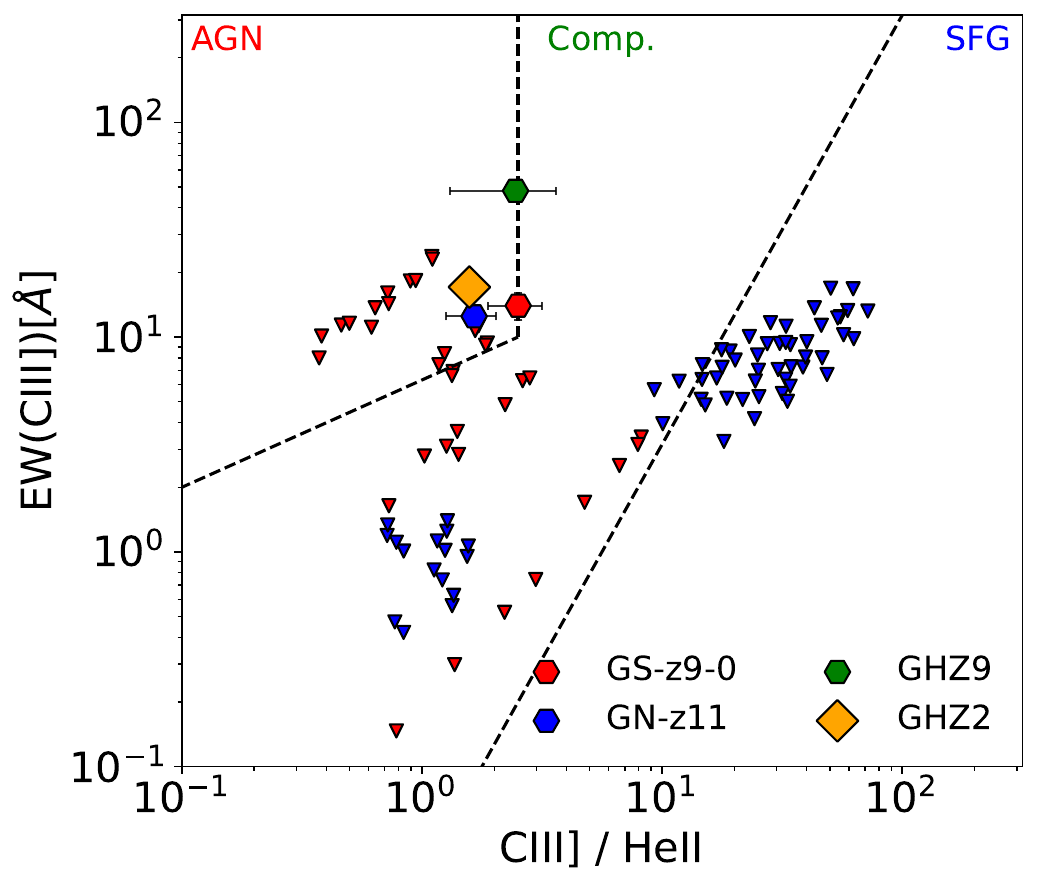}
\includegraphics[trim={0.cm 0.cm 0.cm 0.cm},clip,height=0.2\linewidth,keepaspectratio]{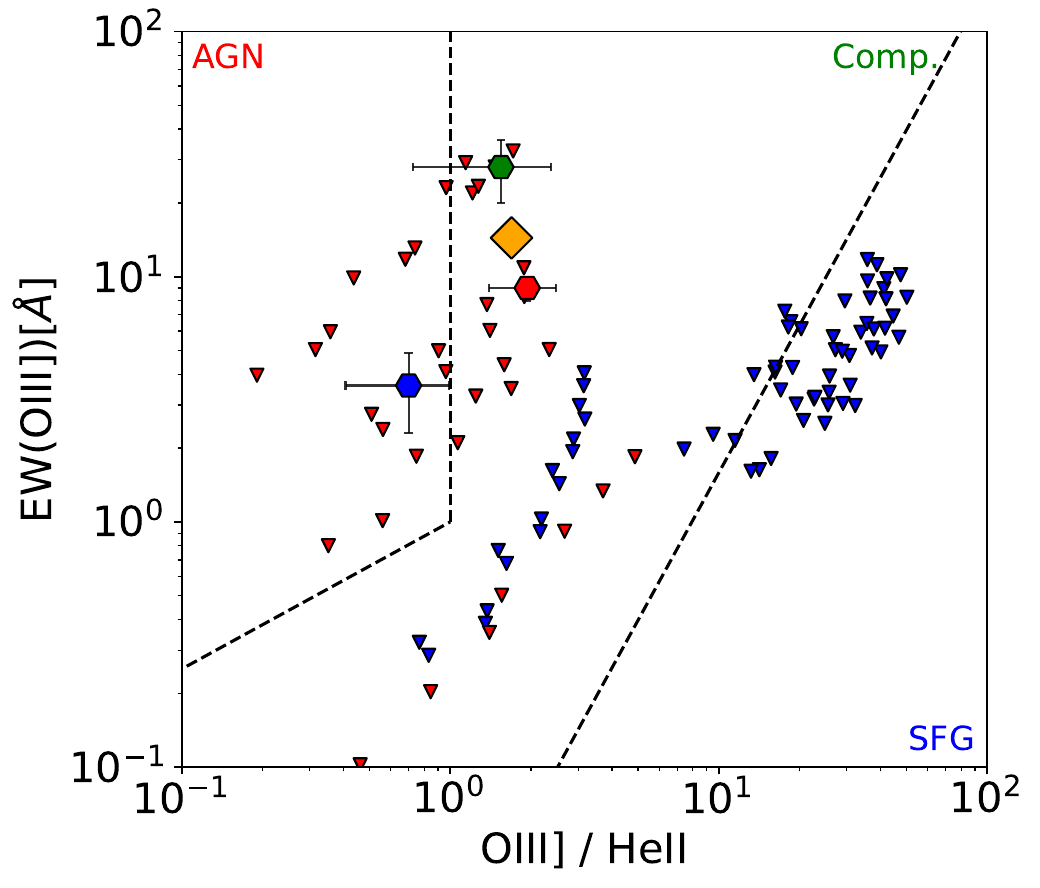}
\includegraphics[trim={0.cm 0.cm 0.cm 0.cm},clip,height=0.2\linewidth,keepaspectratio]{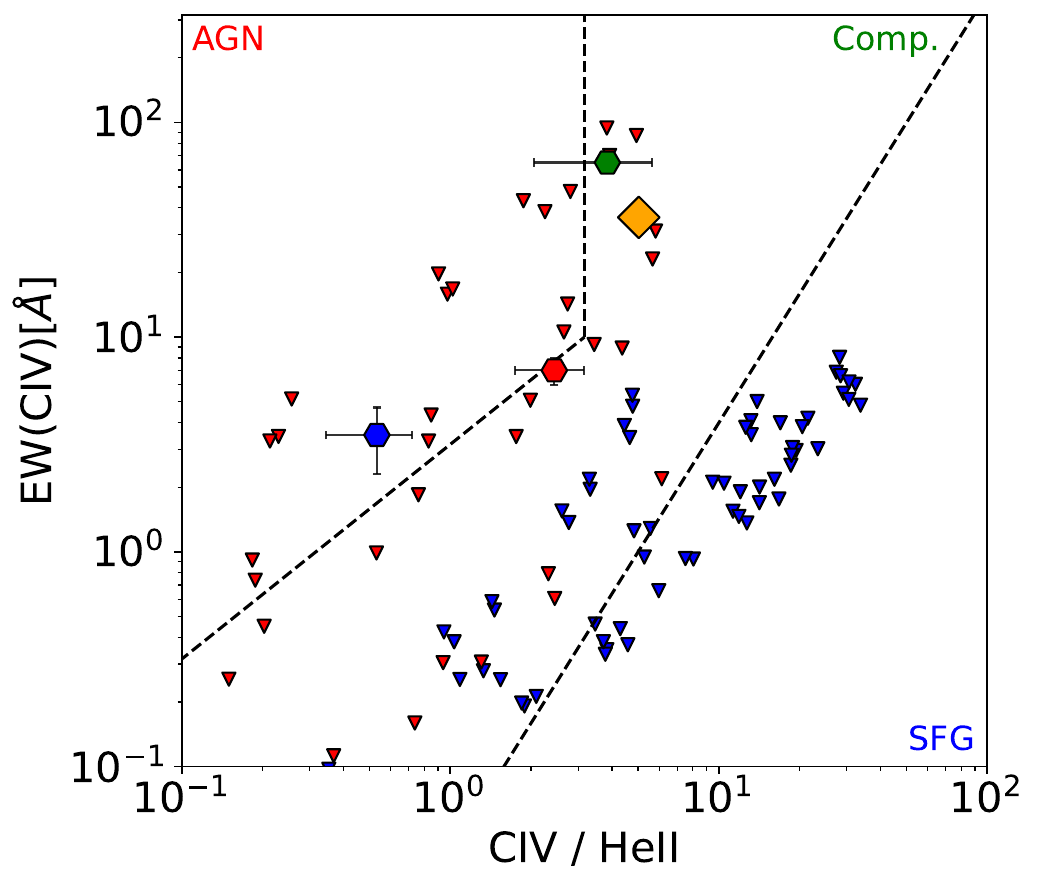}
\includegraphics[trim={0.cm 0.cm 0.cm 0.cm},clip,height=0.2\linewidth,keepaspectratio]{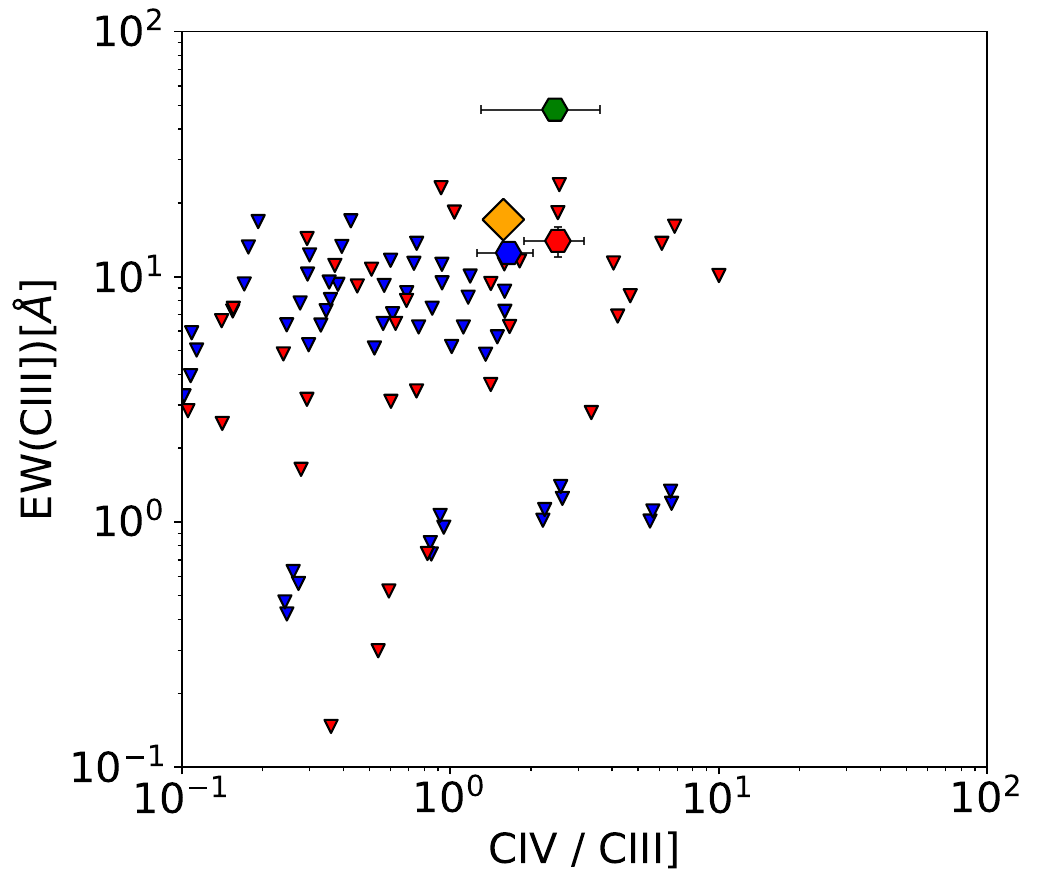}
\includegraphics[trim={0.cm 0.cm 0.cm 0.cm},clip,height=0.2\linewidth,keepaspectratio]{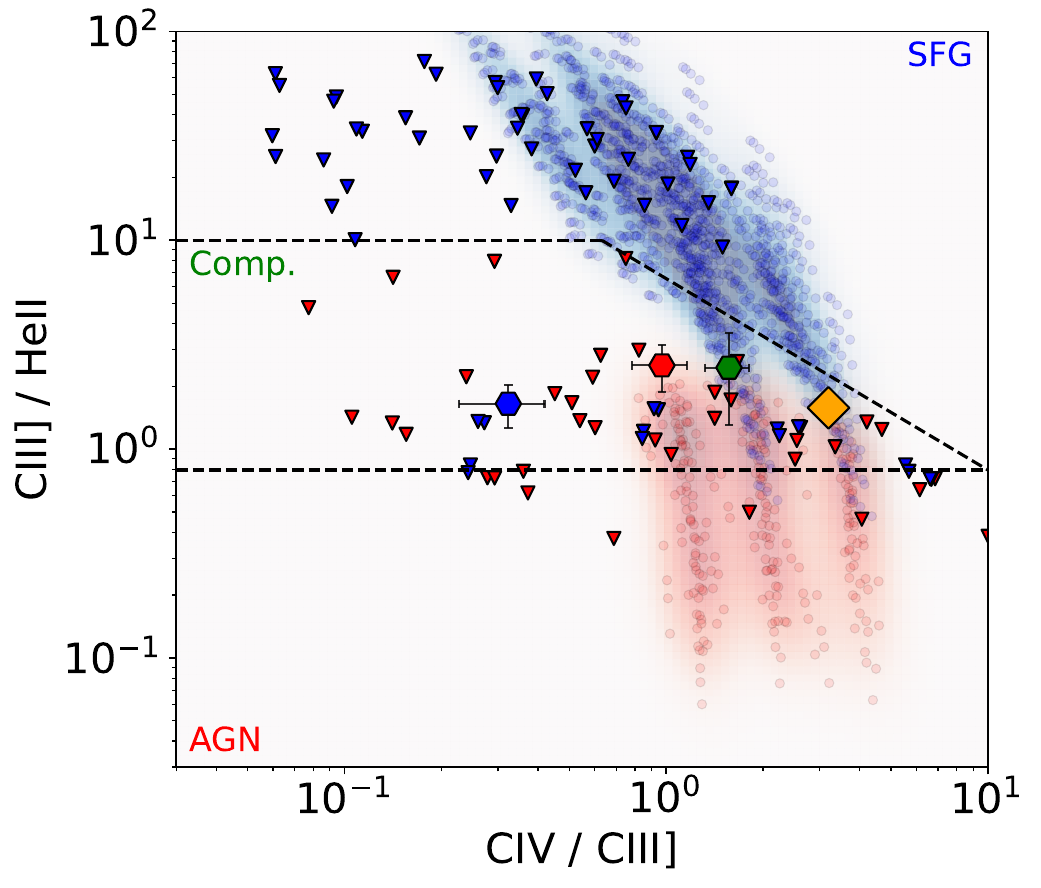}
\includegraphics[trim={0.cm 0.cm 0.cm 0.cm},clip,height=0.2\linewidth,keepaspectratio]{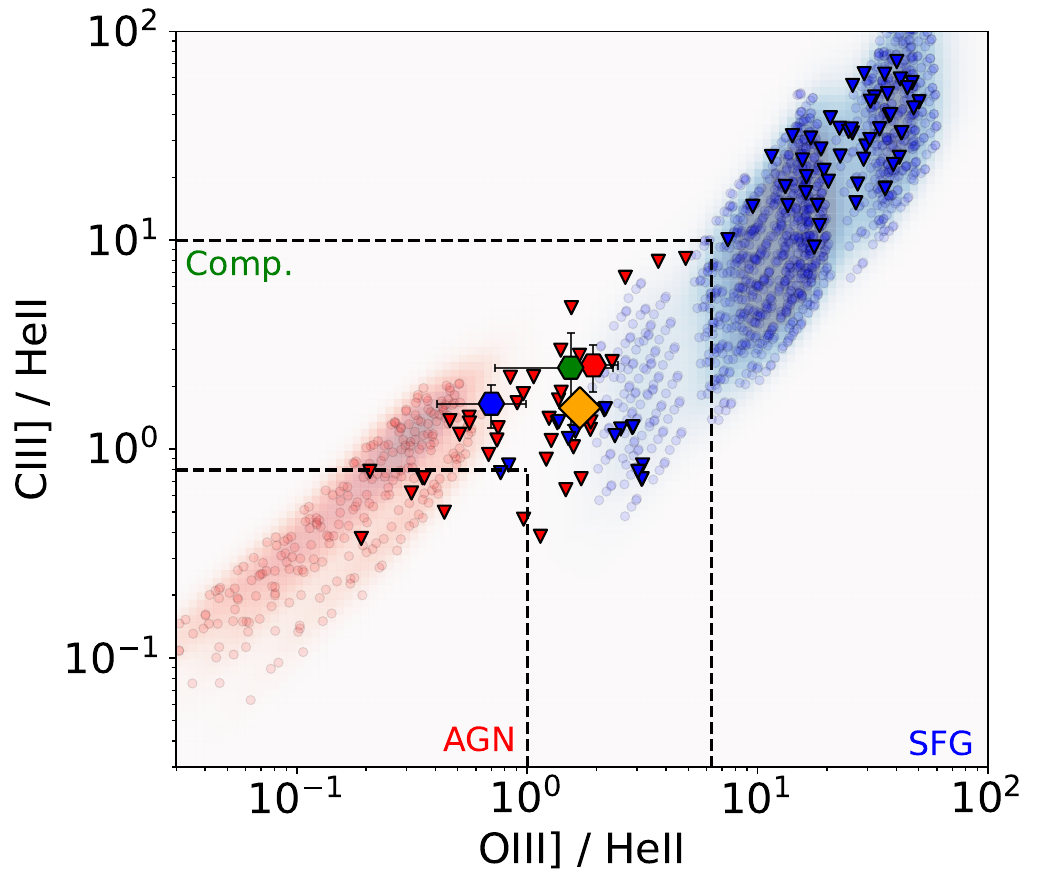}
\includegraphics[trim={0.cm 0.cm 0.cm 0.cm},clip,height=0.2\linewidth,keepaspectratio]{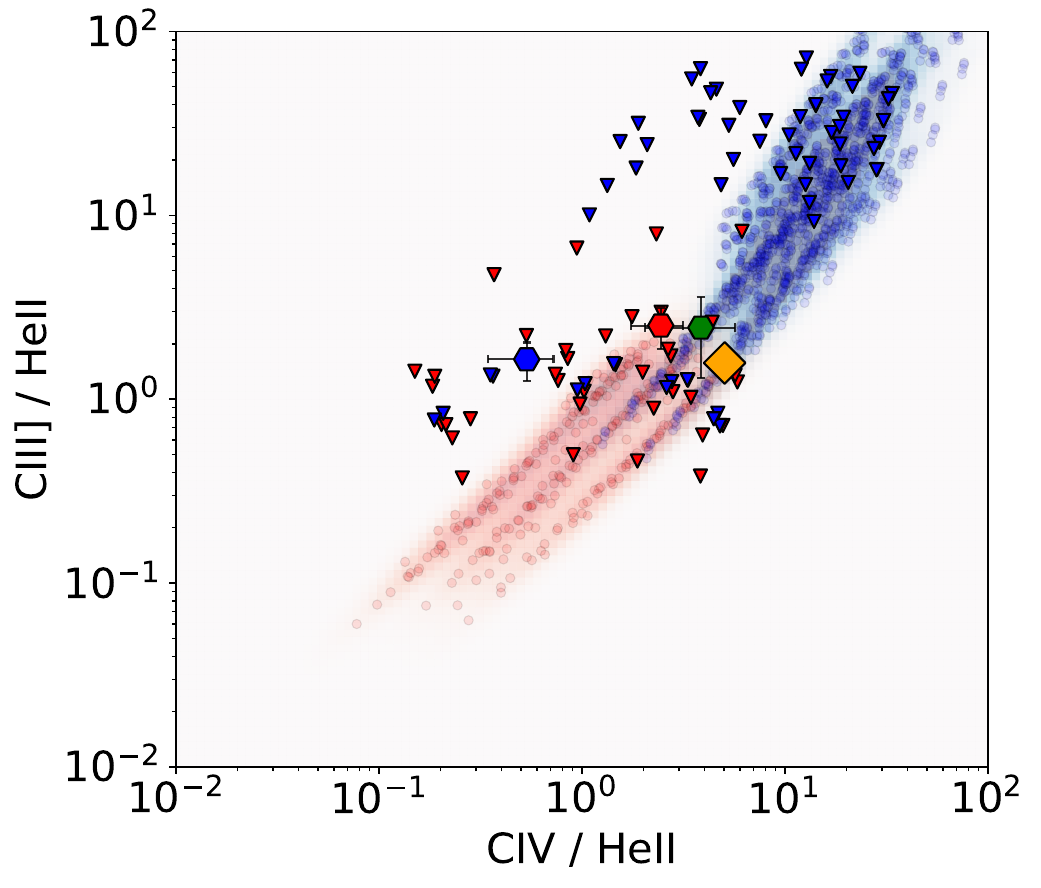}
\includegraphics[trim={0.cm 0.cm 0.cm 0.cm},clip,height=0.2\linewidth,keepaspectratio]{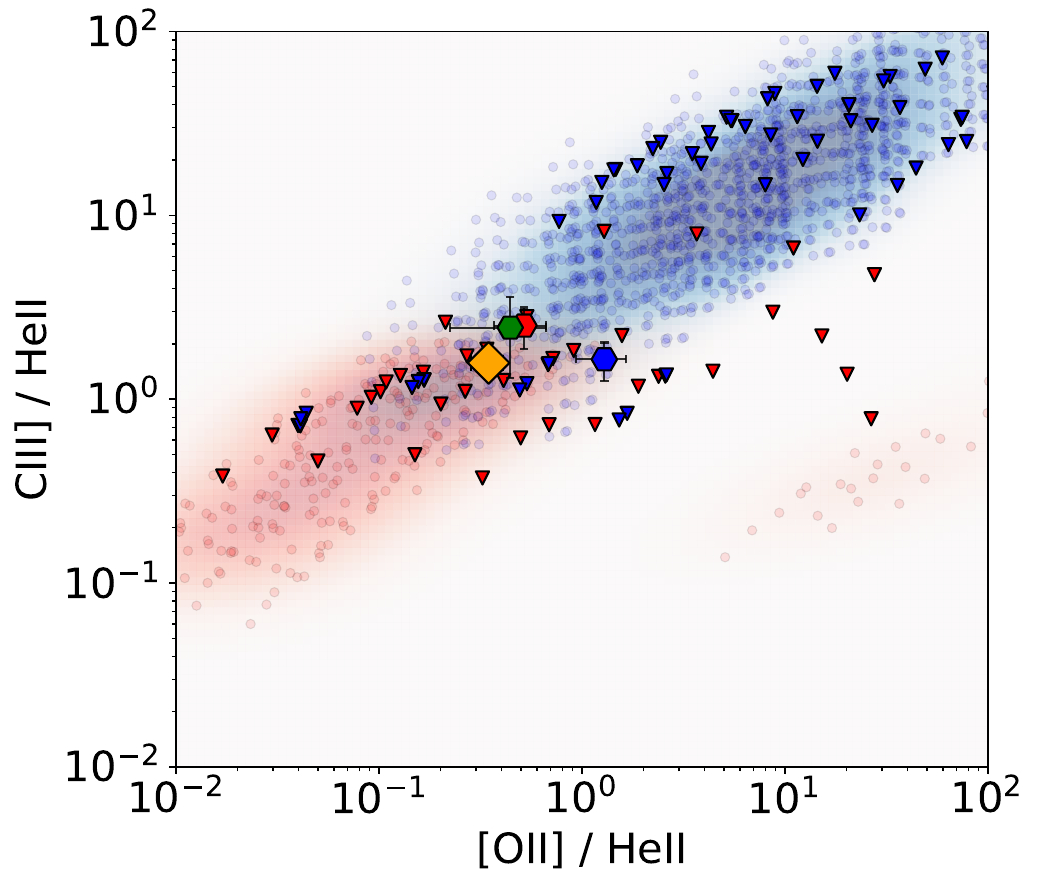}
\includegraphics[trim={0.cm 0.cm 0.cm 0.cm},clip,height=0.2\linewidth,keepaspectratio]{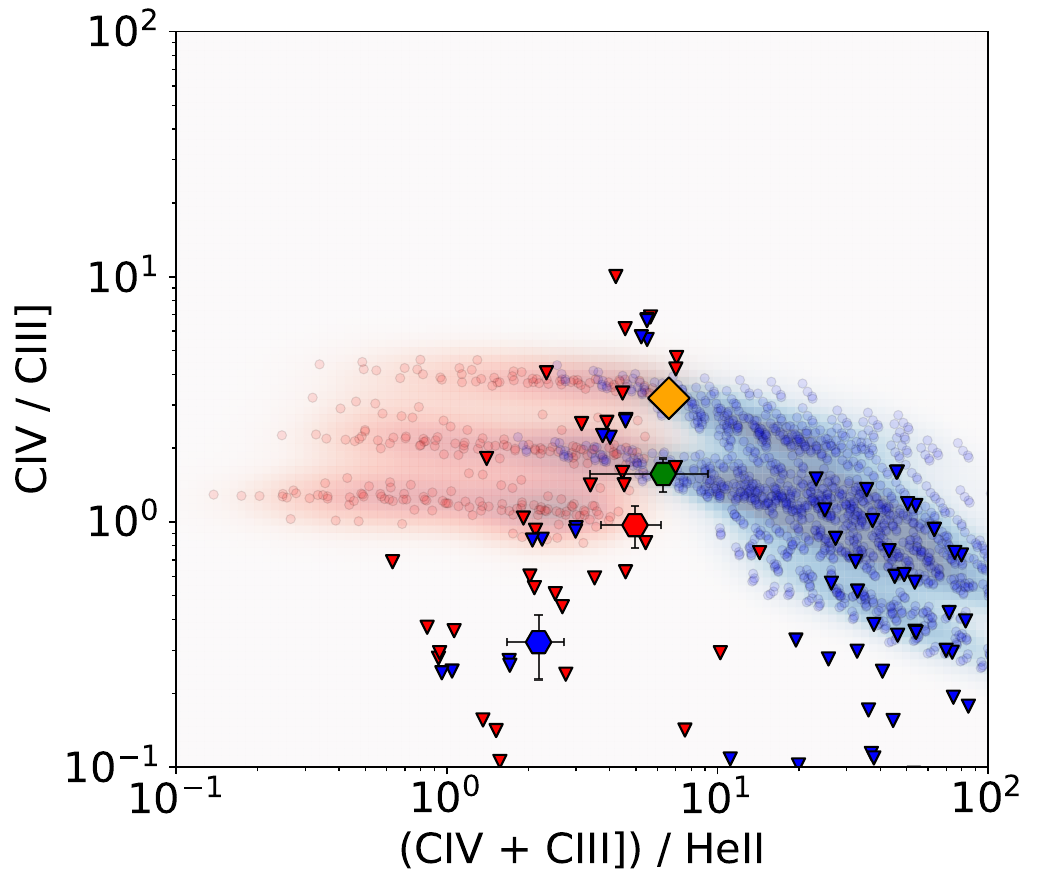}
\includegraphics[trim={0.cm 0.cm 0.cm 0.cm},clip,height=0.2\linewidth,keepaspectratio]{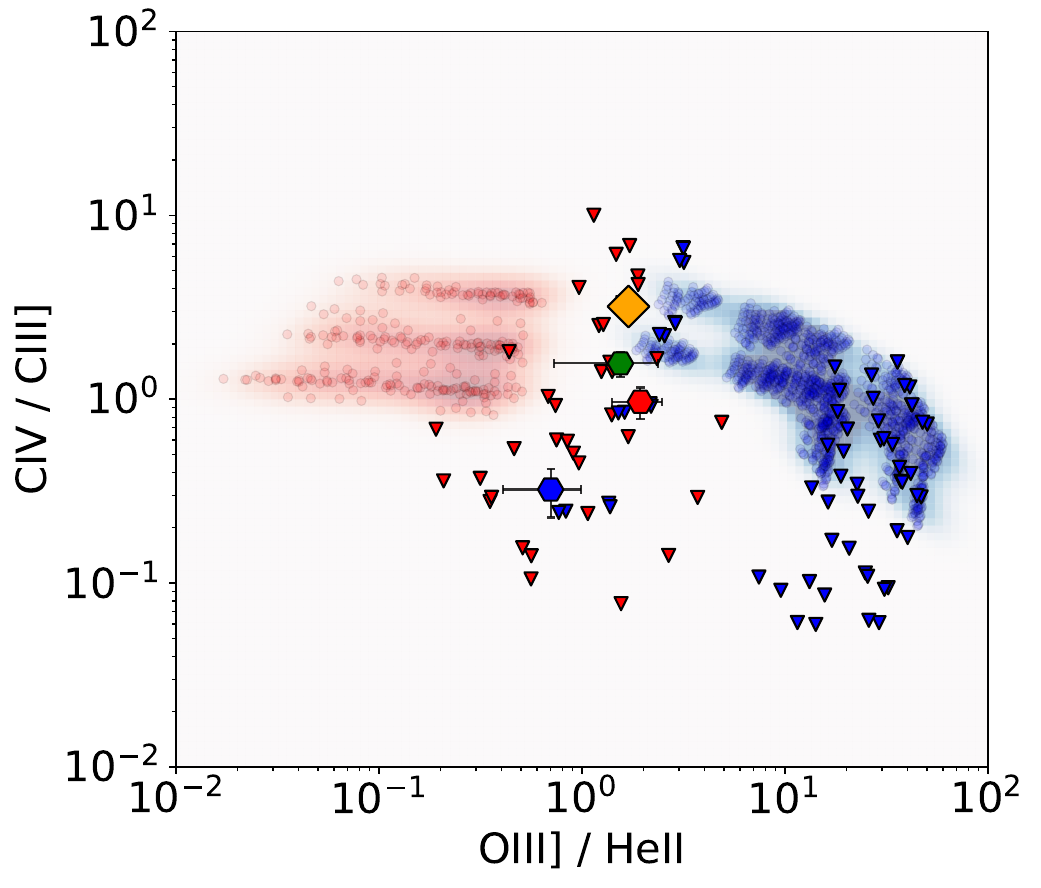}
\includegraphics[trim={0.cm 0.cm 0.cm 0.cm},clip,height=0.2\linewidth,keepaspectratio]{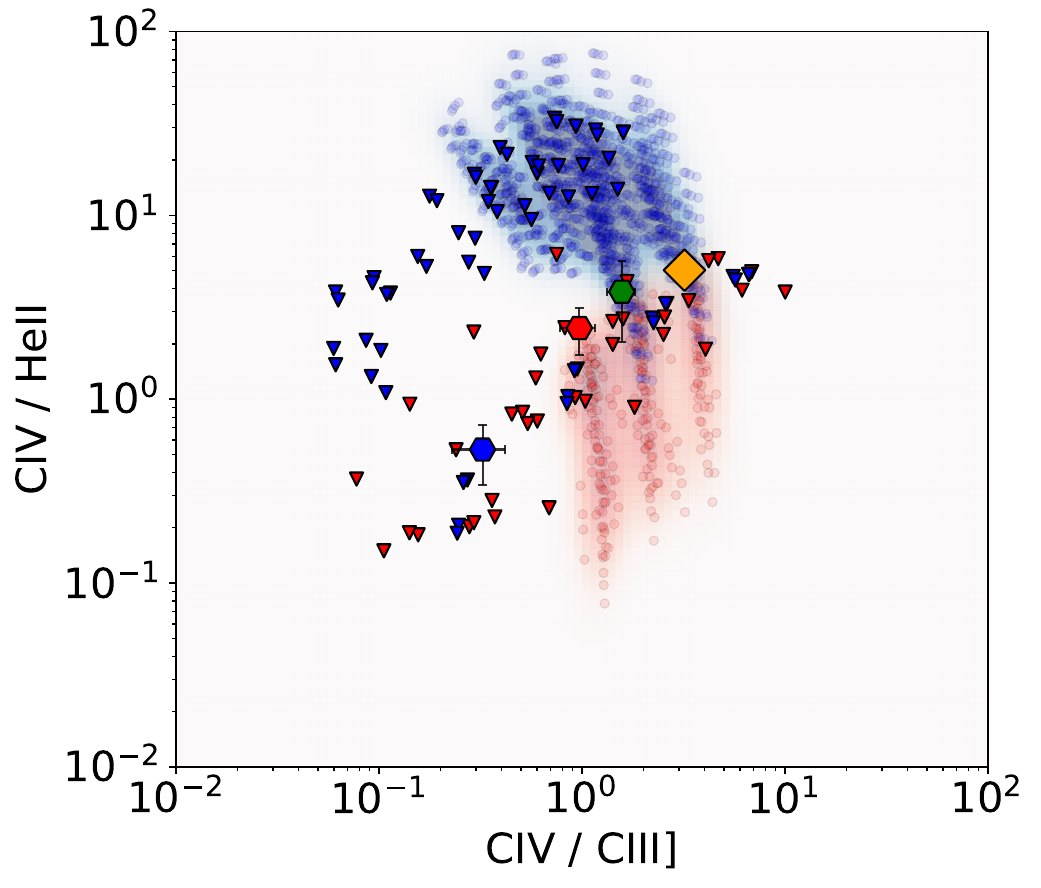}
\includegraphics[trim={0.cm 0.cm 0.cm 0.cm},clip,height=0.2\linewidth,keepaspectratio]{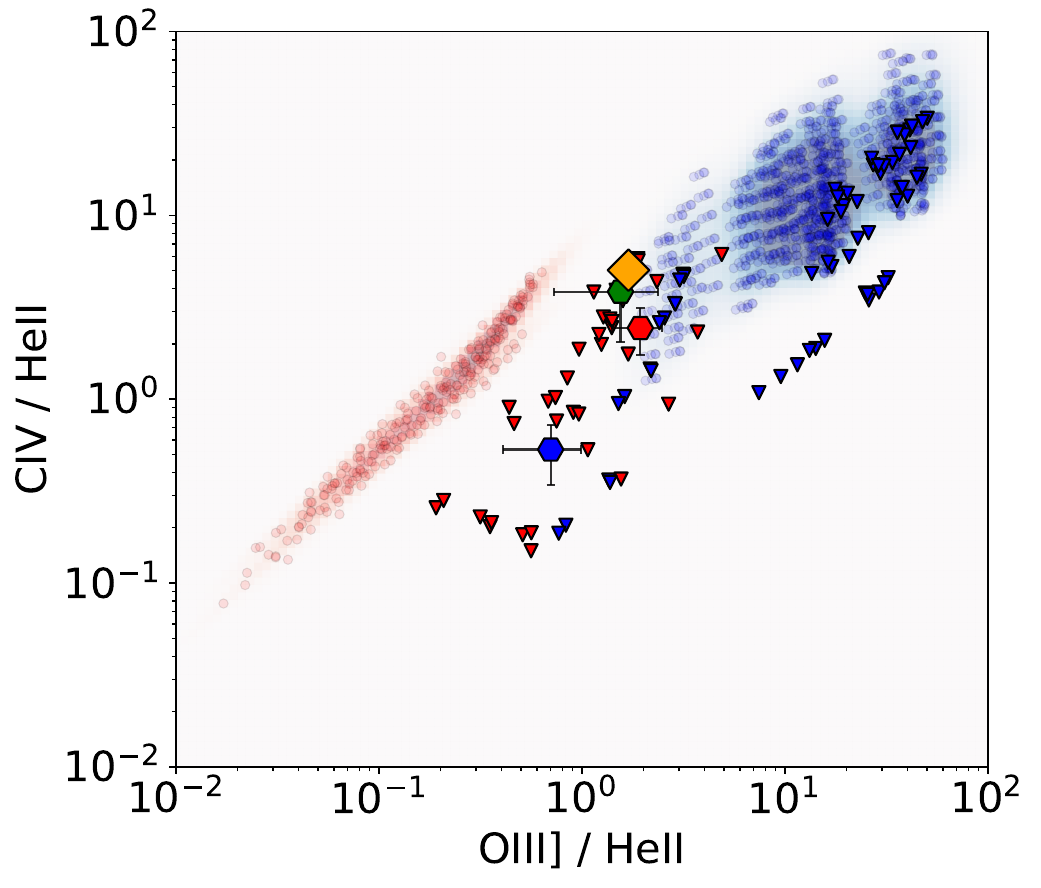}
\caption{The position of GHZ2 (orange-filled diamond with black error bars) in the UV line diagnostic diagrams discriminating between star formation and an AGN as the main ionising source. The AGN and star-forming models by FG16 (NM22) are shown as red and blue circles (triangles), respectively. Dashed demarcation lines are from \citet{Hirschmann2019}. The positions of AGN candidates at z$\gtrsim$9 from the literature are shown when relevant data are available:  
GS-z9-0 \citep[z=9.43;][]{Curti2024}, GHZ9 \citep[z=10.145;][]{Napolitano2024b}, GN-z11 \citep[z=10.6;][]{Bunker2023, Maiolino2023}.} 
\label{fig_DIAGRAMS}
\end{figure*}

We show in Fig.~\ref{fig_lines} snapshots of the final NIRSpec spectrum in regions centered at the position of all emission features discussed in Sect.~\ref{sec:features} together with the relevant fits of the Gaussian profile and of the local UV continuum. The GHZ2 line ratios and EW are compared in Fig.~\ref{fig_DIAGRAMS} to the AGN and star-forming galaxy photoionization models by \cite{Nakajima2022} (NM22 hereafter), and by \cite{Feltre2016} and \cite{Gutkin2016} (FG16 hereafter).

We restrict the comparison to the NM22 grids that are consistent with the physical properties inferred through line scaling relations: -2.8 $\leq$ log\,$U$ $\leq$ -1, and 0.04 $\leq$\zzsun$\leq$ 0.15. In the case of the star-forming models by \cite{Gutkin2016}, we further restrict the parameter space to 0.14 $\leq$ (C/O)/(C/O)$_{\odot}$ $\leq$ 0.72.  To aid interpretation, we include the UV diagnostic demarcation lines from \cite{Hirschmann2019} \citep[see also,][]{Hirschmann2023}, which distinguish between AGN-dominated, star-formation-dominated, and composite sources.

\begin{figure*}[ht]
\centering
\includegraphics[trim={1.6cm 1.95cm 0.55cm 0.8cm},clip,width=0.44\linewidth,keepaspectratio]{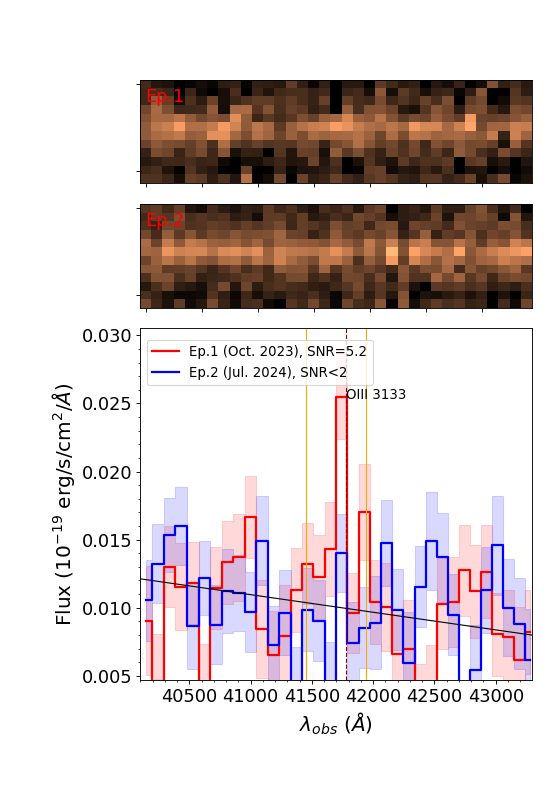}
\includegraphics[trim={1.6cm 5.8cm 0.55cm 6.45cm},clip,width=0.44\linewidth,keepaspectratio]{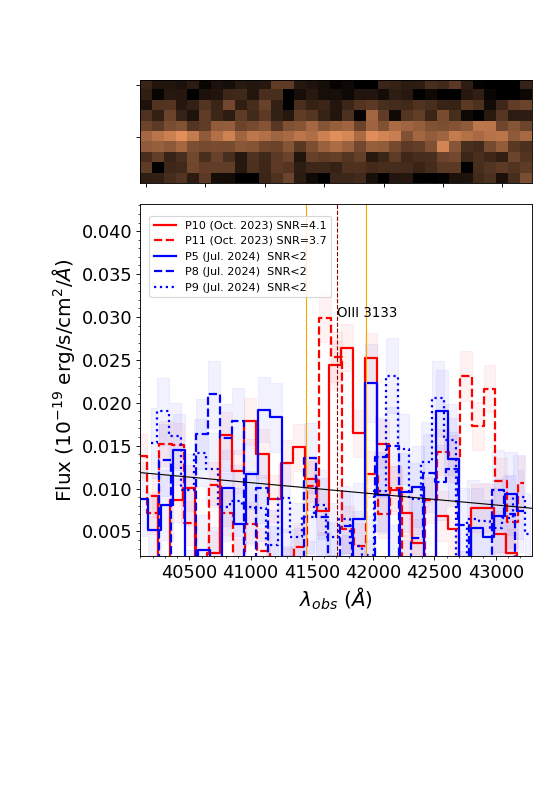}
\caption{\textbf{Left:} Same as Fig.~\ref{fig_OIIIBowen} but for the NIRSpec spectra obtained with an independent data reduction procedure as described in Sect.~\ref{subsec:Bowen}. \textbf{Right:} Same as Fig.~\ref{fig_OIIIBowen} but for single pointing spectra of the different epochs. The red continuous and dashed lines show spectra from the P10 and P11 observations (Oct. 2023), respectively. The blue continuous, dashed and dotted lines show the spectra from P5, P8 and P9 (July 2024), respectively. The relevant 1$\sigma$ uncertainties in each pixel are shown as shaded regions. } 
\label{fig_OIIIBowen_visitsGRB}
\end{figure*}

The region of the \oiiibowen~line of GHZ2 in spectra of the two separated observing epochs of the GO-3073 program obtained by the independent reduction by \citet{Roberts-Borsani2024} \citep[see also,][]{Roberts-Borsani2025b} is shown in the top panel of Fig.~\ref{fig_OIIIBowen_visitsGRB}. The bottom panel of Fig.~\ref{fig_OIIIBowen_visitsGRB} shows the 1D spectra of the five separated pointings comprising our dataset. All the spectra have been normalized to the NIRCam photometry as described in Sect.~\ref{sec:obs}. We find that the line is significantly detected only in the first epoch dataset, and in each of the relevant pointings.

\end{appendix}

\end{document}